\documentclass[prd,superscriptaddress,amsfonts,amssymb,amsmath,showpacs,twocolumn]{revtex4-2}
\usepackage{bm}
\usepackage{amsfonts}
\usepackage{latexsym}
\usepackage[latin1]{inputenc}
\usepackage{graphicx}
\usepackage{amsmath,eqnarray}
\usepackage{palatino}
\usepackage{mathpazo}
\usepackage{textcomp}
\usepackage{float}
\usepackage{booktabs}
\usepackage{dcolumn}
\usepackage{hyperref}
\usepackage{mathtools}
\hypersetup{colorlinks,citecolor=red}

\usepackage{amsmath}
\usepackage{xcolor}
\usepackage{orcidlink}
\usepackage[caption=false]{subfig}
\usepackage{commath}
\def\jnl@style{\it}
\def\aaref@jnl#1{{\jnl@style#1}}

\def\aaref@jnl#1{{\jnl@style#1}}

\def\aj{\aaref@jnl{AJ}}                   
\def\apj{\aaref@jnl{ApJ}}                 
\def\apjl{\aaref@jnl{ApJ}}                
\def\apjs{\aaref@jnl{ApJS}}               
\def\apss{\aaref@jnl{Ap\&SS}}             
\def\aap{\aaref@jnl{A\&A}}                
\def\aapr{\aaref@jnl{A\&A~Rev.}}          
\def\aaps{\aaref@jnl{A\&AS}}              
\def\mnras{\aaref@jnl{Mon.~Not.~Roy.~Astron.~Soc.}}             
\def\prd{\aaref@jnl{Phys.~Rev.~D}}        
\def\prc{\aaref@jnl{Phys.~Rev.~C}}  
\def\prl{\aaref@jnl{Phys.~Rev.~Lett.}}    
\def\qjras{\aaref@jnl{QJRAS}}             
\def\skytel{\aaref@jnl{S\&T}}             
\def\ssr{\aaref@jnl{Space~Sci.~Rev.}}     
\def\zap{\aaref@jnl{ZAp}}                 
\def\nat{\aaref@jnl{Nature}}              
\def\aplett{\aaref@jnl{Astrophys.~Lett.}} 
\def\apspr{\aaref@jnl{Astrophys.~Space~Phys.~Res.}} 
\def\physrep{\aaref@jnl{Phys.~Rep.}}      
\def\physscr{\aaref@jnl{Phys.~Scr}}       
\def\commat{\aaref@jnl{Comm.~Math.~Phys.}}              
\def\science{\aaref@jnl{Science}}               
\def\cqg{\aaref@jnl{Classical Quant.~Grav.}}            
\def\jpcs{\aaref@jnl{JPCS}}                                     
\def\ijmpd{\aaref@jnl{Int.~J.~Mod.~Phys.~D}}                    
\def\grg{\aaref@jnl{Gen.~Relat.~Gravit.}}               
\def\rpp{\aaref@jnl{Rep.~Prog.~Phys.}}          
\def\npa{\aaref@jnl{Nucl.~Phys.~A}}        
\def\lrr{\aaref@jnl{Living Rev.~Rel.}}                   
\def\jcap{\aaref@jnl{J.~Cosmology Astropart.~Phys.}}    
\def\rmp{\aaref@jnl{Rev.~Mod.~Phys.}}   
\def\epjc{\aaref@jnl{Eur.~Phys.~J.~C}}

\allowdisplaybreaks[1]

\begin{document}

\color{black}

\title{A Comprehensive Study of Massive Compact Star Admitting Conformal Motion Under Bardeen Geometry}

\author{Sneha Pradhan\orcidlink{0000-0002-3223-4085}}
\email{snehapradhan2211@gmail.com}
\affiliation{Department of Mathematics, Birla Institute of Technology and
Science-Pilani,\\ Hyderabad Campus, Hyderabad-500078, India.}

\author{P.K. Sahoo\orcidlink{0000-0003-2130-8832}}
\email{pksahoo@hyderabad.bits-pilani.ac.in}
\affiliation{Department of Mathematics, Birla Institute of Technology and
Science-Pilani,\\ Hyderabad Campus, Hyderabad-500078, India.}

\date{\today}

\begin{abstract}

This article primarily investigates the existence of the charged compact star under the conformal motion treatment within the context of $f(Q)$ gravity. We have developed two models by implementing the power-law and linear form of conformal factor, enabling an in-depth comparison in our study. We have selected the MIT Bag model equation of state to describe the connection between pressure and energy density and matched the interior spherically symmetric space-time with the Bardeen space-time. In addition, the present research examines various physically valid characteristics of realistic stars, such as PSR J$1614-2230$, PSR J$1903+327$, Vela X-1, Cen X-3, and SMC X-1. We compare two constructed models by attributing the behavior of density, pressure, equilibrium conditions, and the adiabatic index. We have additionally included a brief analysis of the scenario involving Reissner-Nordstrom spacetime as an external geometry for the matching condition. In contrast to the Reissner-Nordstrom instance, the Bardeen model with the extra term in the asymptotic representations yields a more intriguing and viable result. The current analysis reveals that the resulting compact star solutions are physically acceptable and authentic when considering the presence of charge with conformal motion in $f(Q)$ gravity.

\end{abstract}

\maketitle

\section{Introduction}\label{sec:1}
The study of compact stars has gained significant attention in recent years. Essentially, ``compact stars" encompass celestial objects such as quark stars, white dwarfs, brown dwarfs, and neutron stars. However, compact objects are considered to have small radii and massive masses, and this is the property that gives these objects their intense density. Discovering precise solutions for stellar objects is a major breakthrough in gravitational physics. The utilization of isotropic fluid in creating stellar formations was widely accepted. However, Ruderman \cite{RR} was the first to propose that compact structures have anisotropic characteristics. The analysis of stellar structures has also been conducted by incorporating the equation of state (EoS) in the presence of anisotropic pressure \cite{RB}. Researchers have recently investigated the properties of charged compact stars in the context of an anisotropic fluid for the Her X-I candidate \cite{SK}. Karmarkar formulated a prerequisite condition for space-time that exhibits spherical symmetry and a class I embedding condition \cite{KR}. Additionally, it is noted that the Schwarzchild interior solution leads to the formation of matter according to Karmarkar's condition with pressure isotropy. The study of charged, compact stars is an intriguing field of research for scholars. Rahaman et al. \cite{FR} examined many models of compact stars with electric charge. Singh et al. \cite{KN} provided precise solutions for the structure of anisotropic stars using the Karmakar spacetime.\\
The investigation of conformal symmetries is significant as it enhances our understanding of the intrinsic composition of space-time geometry when addressing the solution of the geodesic equation of motion related to spacetimes. This symmetry greatly facilitates examining the usual connection between geometry and matter. The behavior of the metric is crucial when combined with curves on a manifold in the context of relativity, which will be discussed in the following section. Several intriguing literary works propose that compact stars can be effectively represented by solutions that allow for a single parameter group of conformal motions. Herrera and his colleagues \cite{LH1,LH2,LH3,LH4} were among the early researchers that provided a comprehensive analysis of the spheres that can accommodate a one parameter group of conformal motions. There are also some significant findings employing conformal killing vectors (CKVs) in recent literature \cite{MFS,Frr,ADA,CAk,FR2}. Mak and Harko \cite{MK} derived a precise solution that describes the interior of a charged strange star with spherical symmetry based on the assumption of a one-parameter group of conformal motions. The issue of discovering static, spherically symmetric anisotropic compact star solutions has been investigated in general relativity through conformal motions. Esculpi and Aloma \cite{ME} investigated two novel families of compact star solutions characterized by charged anisotropic fluid capable of accommodating a one-parameter group of conformal motion.

While general relativity (GR) is a remarkable, well-established, and highly relevant theory, numerous valuable enhancements have been proposed by scholars in recent times. Several gravitational theories are encompassed by the following modifications: $f (R)$ \cite{1,2,3,4,5}, $f (R, G)$ \cite{6,7}, $f (G)$ \cite{8}, $f (R, T )$ \cite{9,10}, $f (Q)$ \cite{11}, $f (R, \phi)$ \cite{12}, and $f (R, \phi, X)$ \cite{13}. These theories entail modifications to traditional GR to better explain the accelerated Universe and investigate circumstances for which GR might not be sufficient. This led to the search for altered or expanded gravitational theories that may effectively tackle these issues. The researcher has discovered two enigmatic constituents, namely dark matter (DM) and dark energy (DE), responsible for the accelerated expansion of the Universe. These modified theories of gravity are presumed to provide improved explanations for the issues of dark matter (DM) and dark energy (DE) \cite{14,15,16}. Moreover, these modified theories of gravities are also highly beneficial for examining stellar model formations. In this study, we incorporate the symmetric teleparallel gravity known as $f(Q)$ gravity, in which the non-metricity $Q$ plays a vital role in describing the gravitational interaction between the particles.
The "co-incident gauge," a commonly employed technique in this theory, is utilized to verify that the affine connection is zero, while maintaining the metric as the basic fundamental variable.
This particular gauge selection has continuously been employed in numerous research projects investigating extensions of the Standard Theory of General Relativity (STGR). The $f (Q)$ theory, presented by J. B. Jimenez and colleagues in 2018 \cite{JBJ}, has acquired acceptance among cosmologists. The main difference between $f (Q)$ gravity and classical GR lies in the nature of the affine connection rather than the characteristics of the physical manifold. The authors in the work \cite{JBJ} have shown that $f (Q)$ gravity is equivalent to GR in a flat spacetime. Several astrophysical object models have been explored in this newly developed $f(Q)$ theory \cite{OS1,OS2,SP}. In addition to compact objects, researchers have also investigated the $f(Q)$ gravity in cosmology and have discovered several acceptable cosmological models that can explain the current accelerated expansion of our Universe. In the work \cite{SM1}, we have taken a power law model of $f(Q)$ and examined the cosmological constraints observationally. We use Markov Chain Monte Carlo (MCMC) statistical analysis to compare our model to the Hubble, Pantheon+SHOES, and their combined observational data. The results of our study indicate that our constructed model in $f(Q)$ gravity can explain the observed cosmological parameters and provide a coherent explanation for the Universe's accelerating expansion despite some variations from the $\Lambda\text{CDM}$ model. EoS is a valuable tool for exploring compact object equilibrium structures in GR. Recent research has focused on the physical features of strange quark combinations using the EoS, notably the MIT bag model. A suitable EoS was utilized to determine the accelerating expansion of the early cosmos using a quark bag model \cite{b1}. Deb et al. \cite{a1} used the MIT bag model and Schwarzschild metric as an exterior space-time to solve Einstein's field equations for compact stellar objects without singularities. Coley and Tupper \cite{a2} investigate ideal fluid spherically symmetric spacetimes with proper CKV inheritance. Abbas and Shahzad \cite{a3} investigated a new solution for an isotropic compact star model with conformal motion using Rastall's theory. They analyzed many physical elements of the model to observe compact star behavior. Jape et al. \cite{a4} developed a new type of charged anisotropic exact model with conformal symmetry in static spherical spacetime. The model was tested for its physical acceptability as a realistic stellar model. In this work, our main focus is to examine the development of charged compact stellar objects that accept CKV using two different models of conformal factor. In addition, we employ Bardeen geometry as the external space-time framework and assume isotropic matter for our present research.\\
Our paper is organized as follows: In section-\ref{secII}, we briefly discuss the mathematical formalism of $f(Q)$ gravity and derive the field equation. Next, by introducing the conformal motion technique, we develop two models in our study in sec-\ref{secIII}. Furthermore, we have examined various physically valid characteristics and stability of realistic stars, such as PSR J$1614-2230$, PSR
J$1903+327$, Vela X$-1$, Cen X$-3$, and SMC X$-1$ in sec-\ref{seciv} and sec-\ref{secv}. A brief comparative study has been done between R-N space-time and Bardeen space-time as an outer structure in sec-\ref{vi}. Finally, we conclude in sec-\ref{vii}.

\section{Basic formulation of $f(Q)$ gravity : Field equation}\label{secII}
The action integral for $f(Q)$ gravity can be generalized as : \cite{c1},
\begin{eqnarray}
    \label{1}
&&\hspace{-0.5cm} S=\int \sqrt{-g}\Big[\frac{1}{16 \pi}f(Q)+\mathcal{L}_M\Big] d^4x.
\end{eqnarray}

Here, $\mathcal{L}_M$ is known as Lagrangian matter density, $g$ is the determinant of the metric tensor, and $d^4(x)$ is known as a four-volume element along the 4-dimensional space-time co-ordinate ($t,r,\theta,\phi$).

The following expression describes the non-metricity tensor associated with the affine connection:
\begin{equation}
\label{4}
Q_{\kappa\mu\nu}\equiv \nabla_\kappa g_{\mu\nu}=\partial_{\kappa}\,g_{\mu\nu}-\Gamma^{\delta}{}_{\kappa\mu}\,g_{\delta \nu}-\Gamma^{\delta}{}_{\kappa\nu}\,g_{\mu\delta}.
\end{equation}

where $\Gamma^{\delta}{}_{\kappa\mu}$ denotes the general affine connection, which could be expressed as,

\begin{equation}
    \Gamma^{\delta}{}_{\kappa\mu} = \left\{
\begin{array}{c}
\delta \\ \kappa \mu 
\end{array}
\right\}+K^{\delta}{}_{\kappa \mu}+L^{\delta}{}_{\kappa\mu}.
\end{equation}
In the above expression, $\left\{
\begin{array}{c}
\delta \\ \kappa \mu 
\end{array}
\right\}$, $L^{\delta}{}_{\kappa\mu}$, and $K^{\delta}{}_{\kappa \mu}$ are denoted as Levi-Civita connection, contortion tensor, and disformation tensor, respectively. The geometrical expression for the aforementioned quantities are given by:
\begin{eqnarray}
    \left\{
\begin{array}{c}
\delta \\ \kappa \mu 
\end{array}
\right\} &=& \frac{1}{2}\,g^{\delta\lambda}\left(\partial_{\kappa}g_{\lambda\mu}+\partial_{\mu}g_{\delta\nu}-\partial_{\delta}g_{\kappa\mu}\right),\\
K^{\delta}{}_{\kappa \mu} &=& \frac{1}{2} T^{\delta}{}_{\kappa \mu} + T_(\kappa{\,}^{\delta}{\,}_\mu),\\
L^{\delta}{}_{\kappa\mu} &=& \frac{1}{2} Q^{\delta}{}_{\kappa\,\mu}-Q_(\kappa{\,}^{\delta}{\,}_\mu).
\end{eqnarray}
In the above expression, $T^{\delta}{}_{\kappa \mu}$ denotes torsion tensor. In the STEGR formalism, another significant quantity is the superpotential 
$P_{\,\,\,\,\mu\nu}^{\lambda}$
which is defined as follows:

\begin{equation}
\label{5}
P_{\,\,\,\,\mu\nu}^{\lambda}=-\frac{1}{2}L^\lambda_{\,\,\,\,\mu\nu}+\frac{1}{4}\left(Q^{\lambda}-\tilde{Q^{\lambda}}\right)g_{\mu\nu}-\frac{1}{4}\delta^{\lambda}_{\,\,(\mu\,} Q_{\nu)},
\end{equation}

The following expression determines the trace of the non-metricity tensor:

\begin{equation*}
Q_{\lambda}=Q_{\lambda\,\,\,\,\,\mu}^{\,\,\,\mu}, \quad \quad \tilde{Q}_{\lambda}=Q^{\mu}_{\,\,\,\,\lambda\mu}.
\end{equation*}

At last, the scalar form of non-metricity could be written as,

\begin{equation}
\label{6}
Q=-Q_{\lambda\mu\nu}P^{\lambda\mu\nu}.
\end{equation}

 The field equations for $f(Q)$ theory are derived by varying the action in equation \eqref{1} with respect to the inverse metric tensor $g^{\mu\nu}$ as, \cite{c2}

\begin{eqnarray}
\label{field}
&&\frac{2}{\sqrt{-g}}\nabla_{\lambda}\left(f_{Q}\sqrt{-g}\,P^{\lambda}_{\,\,\,\,\mu\nu}\right)+f_{Q}\left(P_{\mu\lambda\kappa}Q_{\nu}^{\,\,\,\lambda\kappa}-2Q^{\lambda\kappa}_{\,\,\,\,\,\,\,\,\mu}\, P_{\lambda\kappa\nu}\right)\nonumber \\&&\hspace{0.9cm}+\frac{1}{2}f\,g_{\mu\nu}=-8\pi T_{\mu\nu}^{\text{eff}}
\end{eqnarray}

Here, $f_Q = \frac{\partial f(Q)}{\partial Q}$. The formula for energy momentum tensor $T_{\mu\nu}$ could be written as,

\begin{eqnarray}
 T_{\mu\nu} &=& -\frac{2}{\sqrt{-g}}\frac{\delta(\sqrt{-g}\mathcal{L}_m)}{\delta g^{\mu\nu}};
\end{eqnarray}

Moreover, through the utilization of Eq. \eqref{1}, an additional constraint can be deduced, expressed as:

\begin{equation}
    \nabla_{\mu}\nabla_{\nu}(\sqrt{-g}f_{Q}P^{\lambda}{}_{\mu\nu})=0.
\end{equation}

The imposition of curvature-free and torsion-free constraints enables the affine connection in the form as: \cite{c2}

\begin{equation}\label{eq:13}
    \Gamma^{\lambda}{}_{\mu\nu}=\left(\frac{\partial x^{\lambda}}{\partial \xi^{\alpha}}\right) \partial_{\mu}\partial_{\nu} \xi^{\alpha}.
\end{equation}

We can adopt a specific coordinate choice known as the co-incident gauge, where the affine connection $\Gamma^{\lambda}{}_{\mu\nu}=0$. Subsequently, the non-metricity equation undergoes a reduction, leading to:

\begin{equation}
    Q_{\kappa\mu\nu}\equiv \nabla_\kappa g_{\mu\nu}=\partial_{\kappa}\,g_{\mu\nu}.
\end{equation}

As a result, this simplification streamlines the computation process, with the metric serving as the primary variable. However, it is important to highlight that the action loses diffeomorphism invariance, except when considering the Standard General Theory of Relativity (STGR) \cite{c3}. This challenge can be addressed by employing the covariant formulation of $f(Q)$ gravity. Given that the affine connection mentioned in Equation \eqref{eq:13} is entirely inertial, the covariant formulation can be implemented by first defining the affine connection without gravity \cite{c4}.\\
Moreover, the formula for the electromagnetic energy-momentum tensor $\varepsilon_{i j}$ could be written as, 
$$
\varepsilon_{i j}=2\left(F_{i k} F_{j k}-\frac{1}{4} g_{i j} F_{k l} F^{k l}\right),
$$
Furthermore,
$$
F_{i j}=\mathcal{A}_{i, j}-\mathcal{A}_{j, i}.
$$

The following expression defines the electromagnetic field tensor:

\begin{equation}
 F_{i j, k}+F_{k i, j}+F_{j k, i}=0, \\ \label{Er}
 \left(\sqrt{-g} F^{i j}\right)_{, j}=\frac{1}{2} \sqrt{-g} j^i .
\end{equation}

The electromagnetic four potential is given by $\mathcal{A}_i$, and the four current density is indicated by $j^i$. Within the framework of a stationary fluid configuration and under the assumption of spherical symmetry, the only constituent of the four-current density that has a non-zero magnitude is referred to as $j^0$ and is aligned along the radial direction $r$. Therefore, apart from the radial component $F_{01}$ of the electric field, the static and spherically symmetric characteristics of the electric field indicate that all other components of the electromagnetic field tensor become zero. If the condition $F_{01}=-F_{10}$, which indicates antisymmetry, is met, then the equation (\ref{Er}) is fulfilled. The electric field equation can be derived from equation (\ref{Er}) as follows:

\begin{eqnarray}
E(r)=\frac{1}{2 r^2} e^{\lambda(r)+v(r)} \int_0^r \sigma(r) e^{\lambda(r)} r^2 d r=\frac{q(r)}{r^2},
\end{eqnarray}

where, $\sigma(r) = \frac{e^{\frac{-\lambda}{2}}}{4\pi r^2}  (r^2E)^{'}$ denotes the total charge density, while $q(r)$ represents the overall charge of the system.

In the above mathematical formulation, the spherically symmetric space-time  metric has been considered whose form is,

\begin{equation}\label{metric}
    ds^2=-e^{\nu(r)} dt^2+e^{\lambda(r)} dr^2+r^2(d\theta^2+sin^{2}\theta \, d\phi^2).
\end{equation}
In the current study, we consider the isotropic fluid matter to model the dense star whose components are given by $T_{\mu}^{\nu}=(-\rho,p,p,p)$, where $\rho$ and $p$ are the energy density and isotropic pressure of the fluid. Furthermore, we have calculated the non-metricity scalar for the above metric which is of the form, 
\begin{eqnarray}
    Q = - \frac{2 e^{-\lambda(r)}(r \nu'(r)+1)}{r^2}.
\end{eqnarray}
All the aforementioned constraints have been used to derive the field equation, which is given by;

\begin{eqnarray}
    \label{FE1}
    \rho^{\text{eff}}+\frac{q^2}{r^4} &=& \frac{f(Q)}{2}-f_Q\Big[Q+e^{-\lambda}\big(\frac{\lambda^{\prime}}{r}-\frac{1}{r}\big)+\frac{1}{r^2}\Big],\\
    \label{FE2}
    p^{\text{eff}} -\frac{q^2}{r^4} &=& \frac{-f(Q)}{2}+f_Q\Big(Q+\frac{1}{r^2}\Big),\\ 
    \label{FE3}
p^{\text{eff}} + \frac{q^2}{r^4} &=&  \frac{-f(Q)}{2}+f_Q\Big[\frac{Q}{2}-\nonumber\\&&e^{-\lambda}\Big\{\frac{\nu^{\prime \prime}}{2}+\big(\frac{\nu^{\prime}}{4}+\frac{1}{2 r}\big)\big(\nu^{\prime}-\lambda^{\prime}\big)\Big\}\Big].
\end{eqnarray}

\section{conformal motion treatment}\label{secIII}

Besides isometries, there exist other types of motions that are highly useful in the context of four-dimensional Lorentzian metrics, their properties, and their applications to mathematical physics. Conformal motions, or CKVs, represent motions along which the metric tensor of spacetime remains invariant up to a scale factor known as the conformal factor.
The equation governing CKV is expressed as:

\begin{eqnarray}\label{conf}
    \mathcal{E}_{\xi} g_{i j}=\phi g_{i j}.
\end{eqnarray}

The mathematical quantity on the left-hand side is the Lie derivative of the metric tensor $g_{ij}$ concerning the vector field $\xi$. It is important to emphasize that, generally, $\phi$ represents a function that can vary with the radial coordinate $r$ and time $t$ known as the conformal factor. Significantly, while the factor $\phi$ is constant, Eq. (\ref{conf}) results in a homothetic vector (HV), and killing vectors (KV) are formed when $\phi$ equals zero. Therefore, HVs and KVs are particular cases of CKVs. Several studies suggest that solutions that enable a one-parameter group of conformal motions can be used to simulate compact stars.
We obtain the following conformal killing equations by substituting Eq.(\ref{conf}) into the spacetime (\ref{metric}).

\begin{eqnarray}
    \xi^{1} \nu^{\prime} = \phi, \quad \xi^{4} = K, \quad \xi^{1} = \frac{\phi r}{2}, \quad \xi^{1} \lambda^{\prime}+2 \xi_{1}^{1} = \phi,
\end{eqnarray}

which additionally produces the following simultaneous solution:

\begin{eqnarray}\label{cnf}
    e^{\nu}=H^{2} r^{2}, \quad e^{\lambda}=\left(\frac{I}{\phi}\right)^{2}, \quad \xi^{i}=K \delta_{4}^{i}+\left(\frac{r \phi}{2}\right) \delta_{1}^{i},
\end{eqnarray}
where $H, K$ and $I$ are arbitrary constants. The research carried out in this paper is conducted by employing two separate and efficient models of the conformal factor $\phi(r)$. The complete formulation will be presented in the following subsections:
\subsection{Model-I}
In our first proposed model, under the assumption that $\phi$ depends solely on the radial coordinate $r$, we have assumed the functional form of $\phi(r)$ as,
\begin{eqnarray}\label{phi1}
   \phi(r)=I\sqrt{\psi(r)}
\end{eqnarray}
The above power-law form of conformal factor is well established in the context of GR and it is motivated from the work \cite{C1}.
  An EoS characterizes a fluid consisting of quarks, namely the up, down, and strange quarks. Here, we have employed the MIT Bag model EoS to describe the fluid's pressure and energy density relations. The MIT EoS model is represented by the equation $p=\frac{1}{3}(\rho-4\,\beta)$. The parameter $\beta$ is referred to as the Bag constant of units MeV$\text{fm}^{-3}$ \cite{C2}.

Now, using the Eq.(\ref{cnf},\ref{phi1}) and manipulating the field equations, we get the solution of $\psi(r)$ as:
\begin{eqnarray}
   \psi(r)= \frac{1}{3m}(m+\beta r^2)-\frac{nr^2}{6m}+\frac{C}{r^2}.
\end{eqnarray}
Where $C$ is the integrating constant. We have considered the linear model of $f(Q)=m\, Q+n$ which is motivated by some other studies given in the above references.
This study focuses on the case where $C$ is not equal to zero and uses Bardeen geometry as the external space-time framework to examine compact stars. Therefore, we get the subsequent explicit and precise solution that accurately describes the internal geometrical and physical structure of a strange star:
\begin{eqnarray}
    e^{\nu} &=& H^2r^2\,\,\,;\,\,\,\, E^2 = -\frac{m \big(r^2-12 C\big)}{6 r^4}, \nonumber\\ e^{\lambda} &=& \frac{6 m r^2}{-r^4 (n-2 \beta )+2 m \big(r^2+3 C\big)},\nonumber\\
    \rho^{\text{eff}} &=& \frac{2\beta r^4-m(r^2+6C)}{2r^4},\nonumber\\
    p^{\text{eff}} &=& \frac{-6\beta r^4-m\big(r^2+6C\big)}{6r^4}.
\end{eqnarray}
Furthermore, the physical parameters exhibit a central singularity due to employing conformal symmetries. Indeed, this formalism cannot overcome the core singularity in the physical parameters. Nonetheless, the solutions of a core-envelope type model can be explored to represent the envelope portion of a star. Now, we will consider appropriate boundary conditions to ensure compatibility between the solutions of the interior spacetime.

\subsubsection{Boundary and matching condition}
Now, one of the most important parts is determining the values of the constants. For that, we usually match the interior geometry with the outer space-time. In this study, we have matched the interior space-time with the Bardeen exterior space-time, which is given by,
\begin{eqnarray}
    d s^{2} &=& -L(r) d t^{2}+L(r)^{-1} d r^{2}+r^{2} (d \theta^{2}+ \sin ^{2} \theta d\phi^{2}),~~~
\end{eqnarray}
where $L(r)=1-\frac{2 \mathcal{M} r^{2}}{\left(\mathcal{Q}^{2}+r^{2}\right)^{\frac{3}{2}}}$. Here, $\mathcal{M}$ is the star's total mass, and $\mathcal{Q}$ is the total charge surroundings of the outer region of the star. By applying binomial expansion, one can get the expression of $L(r)=1-\frac{2 \mathcal{M}}{r}+\frac{3 \mathcal{M} \mathcal{Q}^{2}}{r^{3}}+O\left(\frac{1}{r^{5}}\right)$. Here in this expression of $L(r)$, the presence of the fraction term $\frac{1}{r^3}$ distinguishes Bardeen geometry from the usual Reissner-Nordstrom space-time configuration. We shall ignore the term $O\left(\frac{1}{r^{5}}\right)$ and its subsequent quantity because of its modest value. By applying the continuity equation, we have matched the exterior and interior space-time metric potentials at the boundary.
\begin{eqnarray}
    1-\frac{2 \mathcal{M}}{r_{b}}+\frac{3 \mathcal{M} \mathcal{Q}^{2}}{r_{b}{ }^{3}} &=& H^2\,r_{b}^2,\\
    \left(1-\frac{2 \mathcal{M}}{r_{b}}+\frac{3 \mathcal{M} \mathcal{Q}^{2}}{r_{b}{ }^{3}}\right)^{-1} &=& \frac{6 m r_b^2}{2 m \big(r_b^2+3 C-r_b^4 (n-2 \beta )\big)}.\nonumber
\end{eqnarray}
By imposing the above matching conditions, we have derived the values of the following constants,
\begin{eqnarray}
    H &=&\pm \frac{\sqrt{5 m \mathcal{M}-2 \mathcal{M} n r_b^2+4 \beta  \mathcal{M} r_b^2+4 \mathcal{M}-2 r_b}}{\sqrt{12 m \mathcal{M} r_b^2-2 r_b^3}},\\
    C &=& \frac{r_b^2 \left(-3 m^2 \mathcal{M}-12 m \mathcal{M}+4 m r_b+n r_b^3-2 \beta  r_b^3\right)}{6 m (r_b-6 m \mathcal{M})}.
\end{eqnarray}
We have given the numerical values of the constants for model I and model II by varying the model parameter $m,n$ in the table-\ref{table1}. The corresponding numerical values of the constants have been calculated by considering the observational data for the star PSR J1614-2230.

\subsection{Model-II }

Here, we have implemented an alternate model of conformal factor $\phi(r)$ to analyze the compact star, which is given by,
\begin{eqnarray}
    \phi(r)=H+N\,r.
\end{eqnarray}
where $H,N$ are arbitrary constant.
This linear form of conformal factor has been widely studied in the literature \cite{C3, C4}. Our work is well motivated by these articles. By imposing the above linear form of the conformal vector into the motion equations, we get the solution of the field equation as,
\begin{eqnarray}
    e^{\nu(r)} &=& H^2r^2,\nonumber \\
    e^{\lambda(r)} &=& \big(\frac{I}{H+N\,r}\big),\nonumber\\
    \rho^{\text{eff}} &=& -\frac{3 H^2 m+3 H N m r-2 \beta  I^2 r^2}{2 I^2 r^2},\nonumber\\
    p^{\text{eff}} &=& -\frac{H^2 m+H N m r+2 \beta  I^2 r^2}{2 I^2 r^2},\nonumber\\
    E^2 &=&\frac{1}{2I^2r^2}\big(5 H^2 m+11 H N m r-2 I^2 m+\nonumber\\&&I^2 n r^2-2 \beta  I^2 r^2+6 N^2 m r^2\big). \nonumber   
\end{eqnarray}
Upon examining the solution of the field equation, it becomes apparent that there are three constants: H, I, and N. To establish the extra requirement for model II, we will employ the second fundamental condition of the continuity equation, which states that $ p(r=r_b)$ must equal zero, where $r_b$ corresponds to the stellar radius. The derived constants for model-II are given by:
\begin{eqnarray}
       H &=& \mp \frac{\sqrt{6 m \mathcal{M}-3 \mathcal{M} n r_{b}^2+4 \mathcal{M}-2 r_{b}}}{\sqrt{18 m \mathcal{M} r_{b}^2-2 r_{b}^3}}\\
        I &=& -\frac{m \left(6 m \mathcal{M}-3 \mathcal{M} n r_{b}^2+4 \mathcal{M}-2 r_{b}\right)}{4 \beta  r_{b}^3 (9 m \mathcal{M}-r_{b})} \\
       N &=& \pm \frac{\sqrt{2 \mathcal{M} \left(6 m-3 n r_{b}^2+4\right)-4 r_{b}}\times\mathcal{F}}{8 \beta  r_{b} \left(-r_{b}^2 (r_{b}-9 m \mathcal{M})\right)^{3/2}}\\
        H &=& \pm \frac{\sqrt{6 m \mathcal{M}-3 \mathcal{M} n r_{b}^2+4 \mathcal{M}-2 r_{b}}}{\sqrt{18 m M r_{b}^2-2 r_{b}^3}}\\
        I &=& +\frac{m \left(6 m \mathcal{M}-3 \mathcal{M} n r_{b}^2+4 \mathcal{M}-2 r_{b}\right)}{4 \beta  r_{b}^3 (9 m \mathcal{M}-r_{b})} \\
       N &=& \mp \frac{\sqrt{2 \mathcal{M} \left(6 m-3 n r_{b}^2+4\right)-4 r_{b}}\times\mathcal{F}}{8 \beta  r_{b} \left(-r_{b}^2 (r_{b}-9 m \mathcal{M})\right)^{3/2}}
\end{eqnarray}
where $\mathcal{F}=\big(6 m^2 \mathcal{M}+m \mathcal{M} \left(4-3 r_{b}^2 (n-12 \beta)\right) -2 m r_{b}-4 \beta  r_{b}^3\big)$.
In the next section, we will analyze and compare our two constructed models physically.
\begin{table}[t]
 \caption{The corresponding numerical values of the constants for different model parameters where we have taken the observational mass-radius data for the star PSR J1614-2230.}\label{table1}
 \centering
  \begin{tabular}{@{}ccccccccccccc@{}}
            \hline\hline
             &  & Model-I & \\
            \hline
             $m$ & $n$ & H & $C$\\
             \hline
            2 & 0.02 & $0.189272$ & $0.0387991$\\
            3 & 0.05 & $0.701516$ & $0.0208653$ \\
            0.2 & 0.1 & $1427.6$ & $0.21629$\\
            -0.5 & 0.4 & $-802.576$ & $0.255184$\\
            \hline\hline
             &  & \,\,\,\,\,\,\,\, Model-II&  &\\ \hline\hline
            m & n & H & I & N \\
            \hline
            0.5 & -1 & $1.25289$ & $479.831$ & $-601.298$\\
            1.5 & -2 & $0.551493$ & $278.909$ & $-153.87$\\
            -0.5 & 1.5 & $0.532217 $ & $-86.584$ & $46.0298$\\
            -2 & 4 & $0.530841$ & $-344.547$ & $182.848 $\\
             \hline
        \end{tabular}
\end{table}

\section{Physical analysis}\label{seciv}
To attain a well-behaved and feasible solution, the following conditions must be satisfied for a stellar configuration:

\begin{enumerate}
    \item \textbf{Metric potential}
Since the metric potential inherently incorporates the geometric and causal structure of space-time, it must have finite and bounded values. They should not exhibit any singularities within the star or at its boundary, defined as $0 \leq r \leq r_{b}$. From the left panel of Fig.(\ref{metric}) one can observe that, for model I, the metric potential gives the finite values at every point through the stellar region. Furthermore, it demonstrates a continuous and increasing behavior towards the boundary region. However, in the case of model II, the metric potential does display a central singularity. Indeed, this is the one disadvantage of utilizing conformal symmetries by employing the conformal factor $\phi(r)$ as a linear function of the radial coordinate. Except at the star's core, the metric function gives finite and bounded values throughout the stellar configuration.  

\begin{figure*}[htbp]
    \includegraphics[width=8cm, height=5cm]{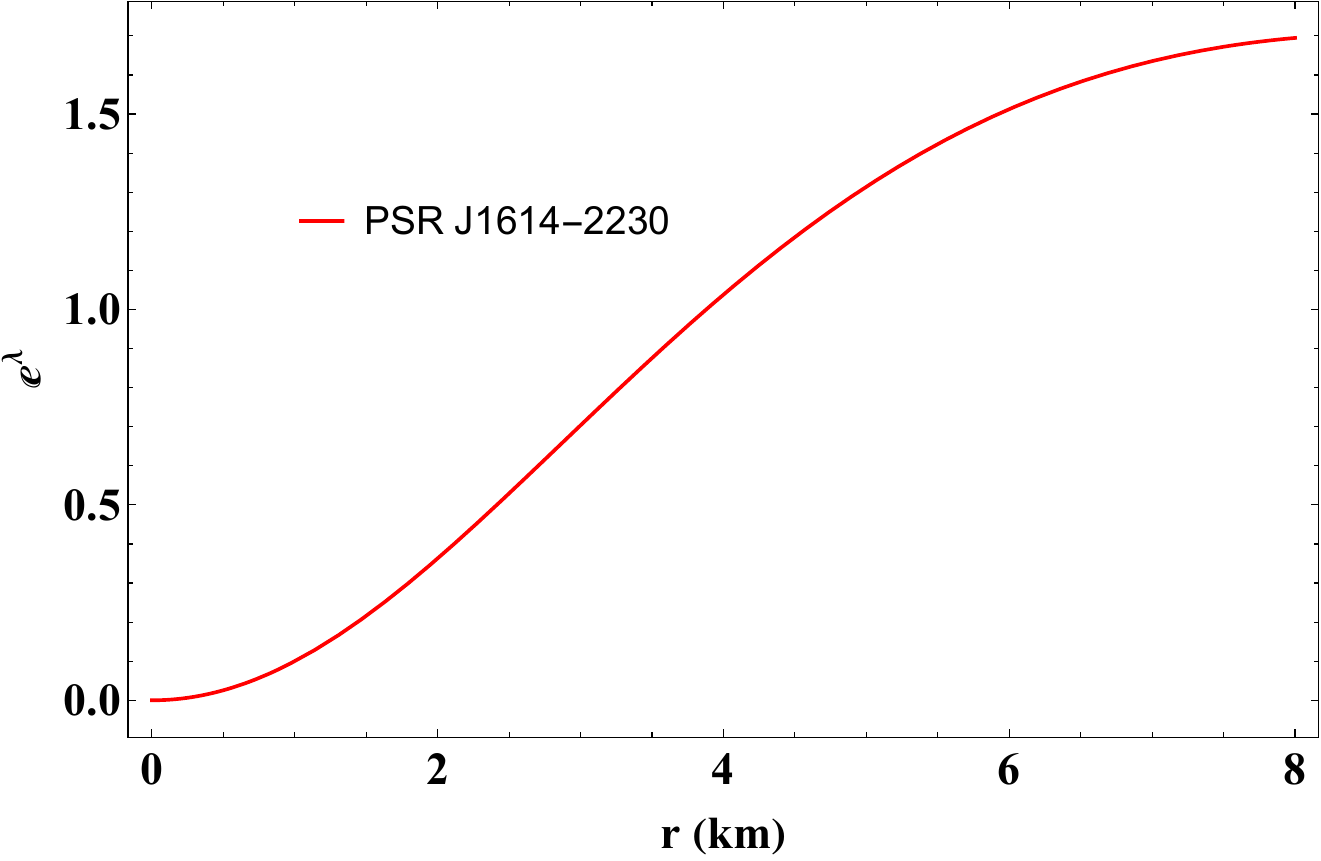}
    \includegraphics[width=8cm, height=5cm]{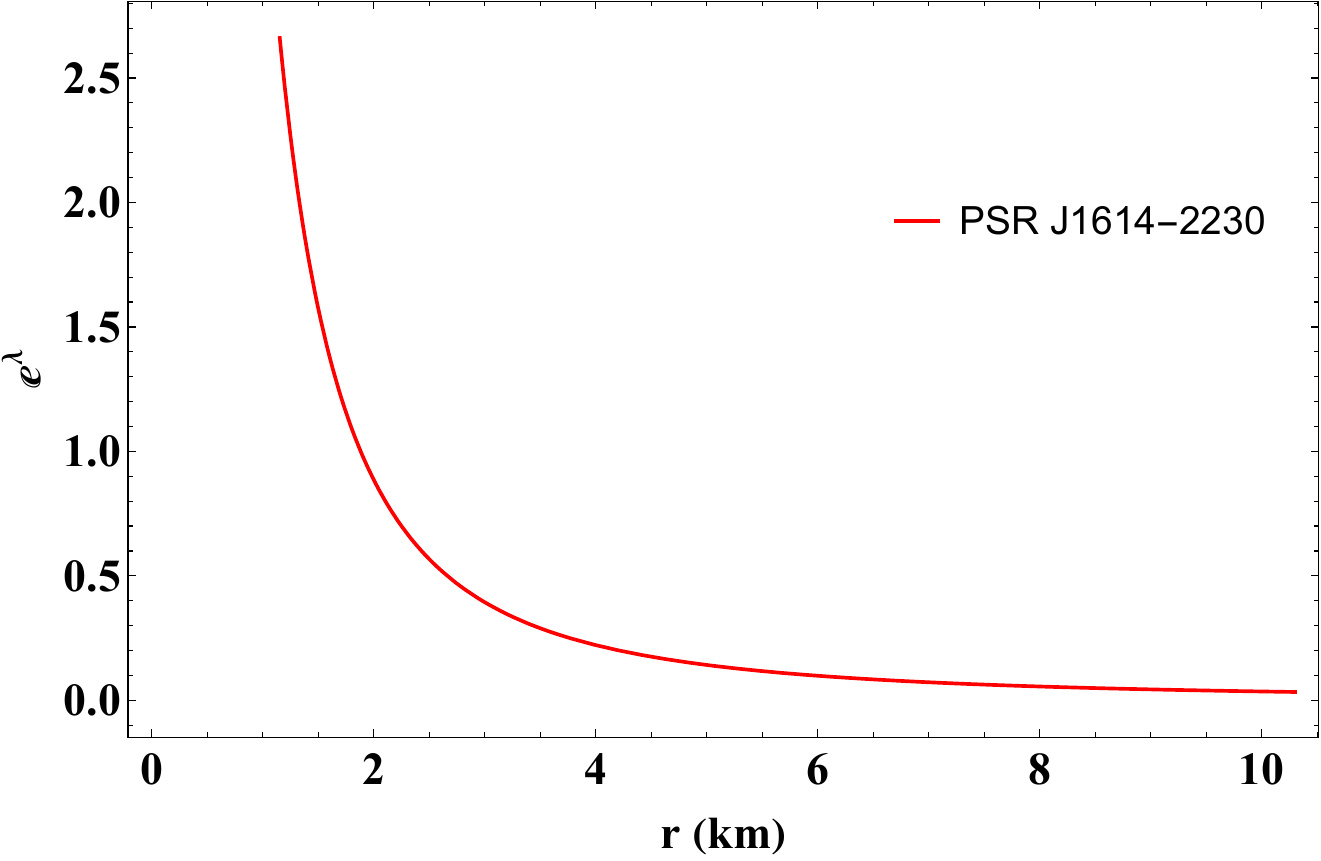}
      \caption{The metric coefficients for model-I(left panel) and model-II (right panel). Here, we consider $m=-2,n=0.02$ for model I and $m=2,n=-1$ for model-II.\label{metric}}
\end{figure*}

\item \textbf{Nature of physical quantities :} 
    This subsection focuses on examining and analyzing the crucial behaviors of physical quantities, specifically pressure and matter density. The following physical properties must be held for a viable and well-behaved stellar model.
    \begin{itemize}
        \item The surface pressure is preciously zero, denoted by $p^{\text{eff}}(r=r_b)=0$. 
        \item The pressure and density functions must exhibit positive values and demonstrate a consistent decrease. Additionally, at the center, their values should reach a maximum. i.e.
        $$
        \begin{aligned}
        & p^{\text{eff}}(0)>0,\left.\quad \frac{d p^{\text{eff}}}{d r}\right|_{r=0}=0,\left.\quad \frac{d^{2} p^{\text{eff}}}{d r^{2}}\right|_{r=0}<0, \quad \rho^{\text{eff}}(0)>0 \\
        & \left.\frac{d \rho^{\text{eff}}}{d r}\right|_{r=0}=0,\left.\quad \frac{d^{2} \rho^{\text{eff}}}{d r^{2}}\right|_{r=0}<0.
        \end{aligned}
        $$
            \end{itemize}
We have conducted a thorough examination and analysis of the physical parameters of many compact stars, such as PSR J1614-2230, PSR J1903+327, Vela X-1, Cen X-3, and SMC X-1, by utilizing different observational data on their mass and radius.  By referring to Fig.(\ref{pressure2}), it is evident that the pressure and matter density of the star exhibits a prominent peak at the central area, followed by a continuous decrease towards the surface region of each compact star. Furthermore, the surface pressure of the star reaches a value of zero near its boundary. The upward concave expansion of the curve for energy density and pressure can be attributed to the dynamics of conformal symmetry and the existence of electric charge.
\begin{figure*}[htbp]
    \includegraphics[width=8cm, height=5cm]{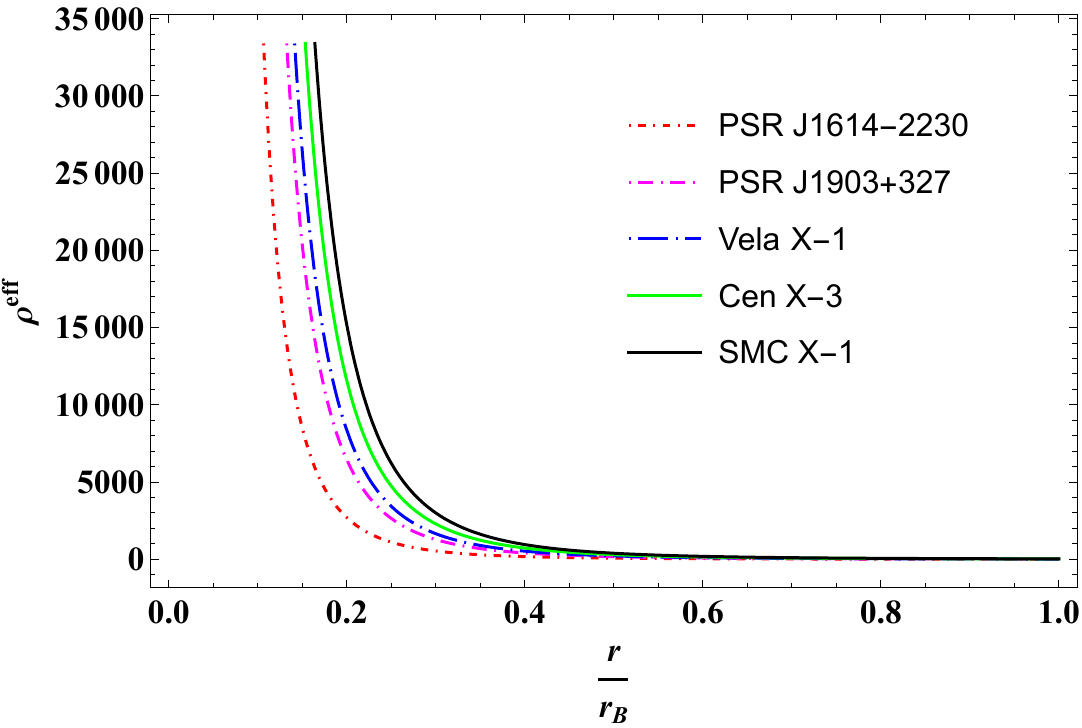}
    \includegraphics[width=8cm, height=5cm]{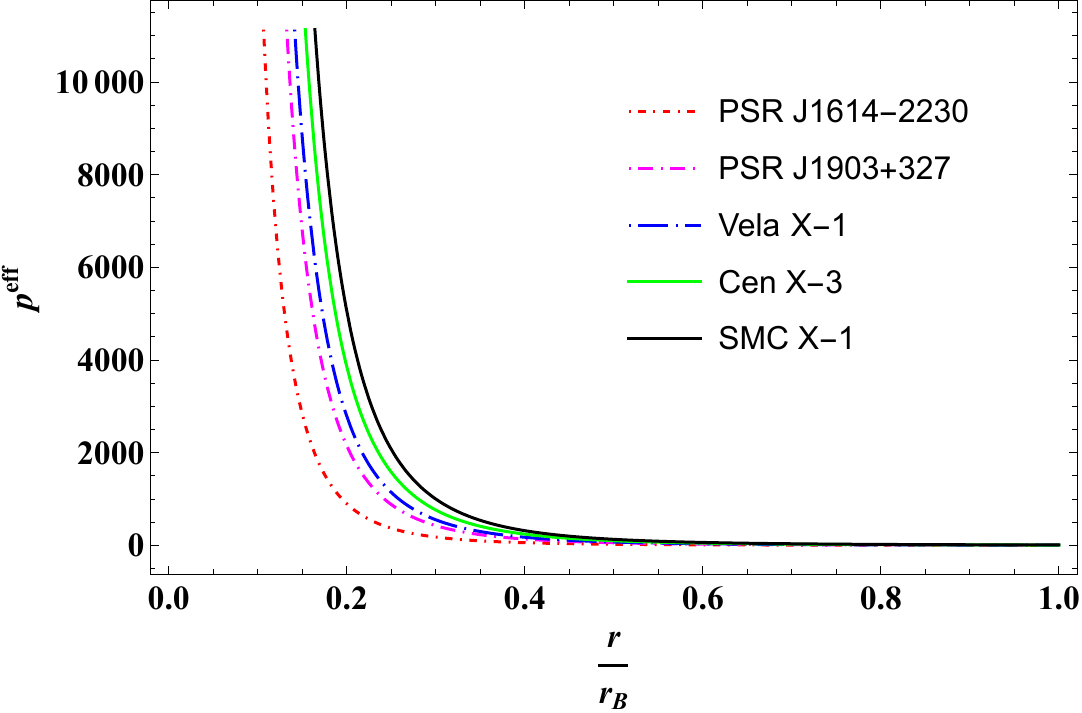}
    \includegraphics[width=8cm, height=5cm]{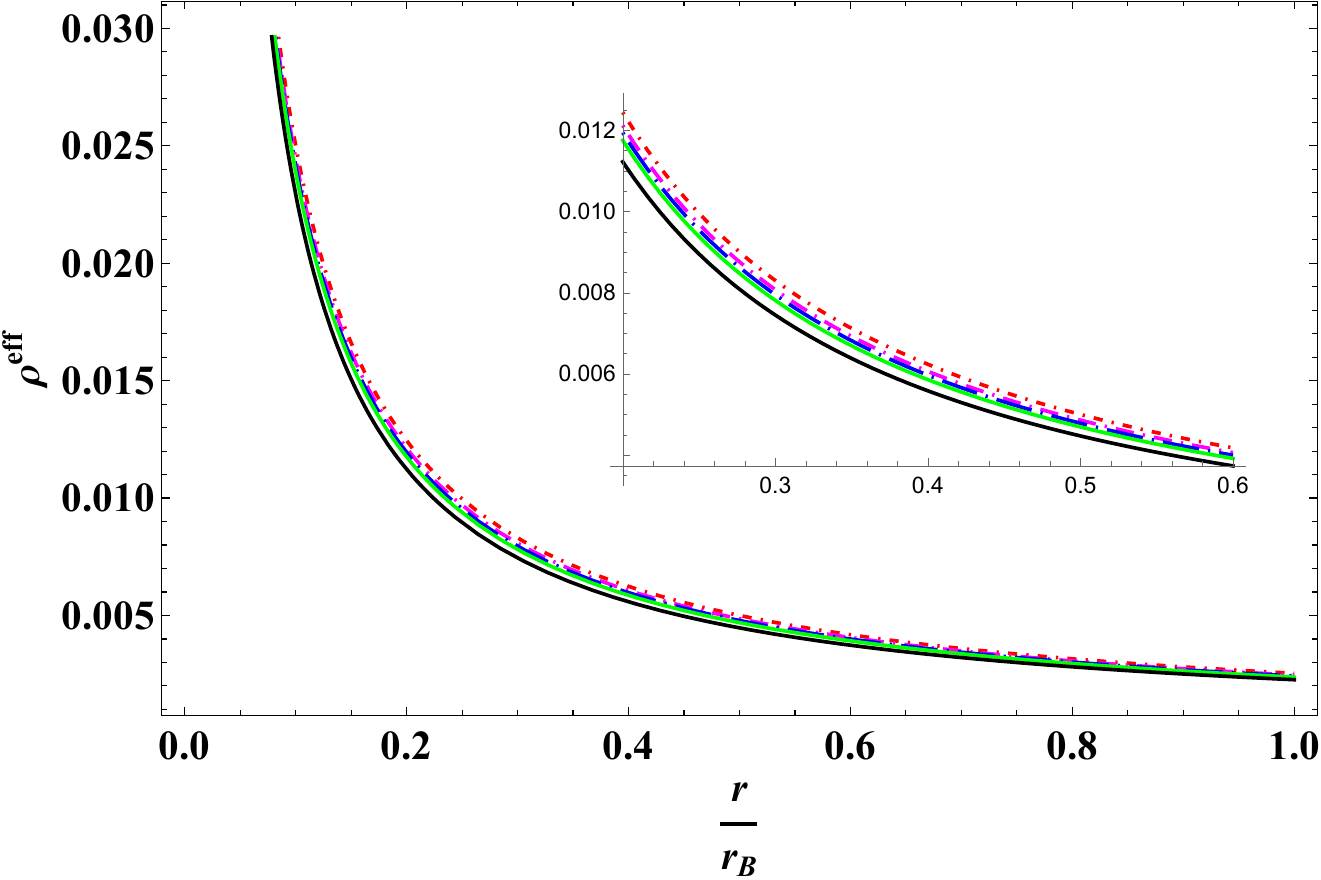}
    \includegraphics[width=8cm, height=5cm]{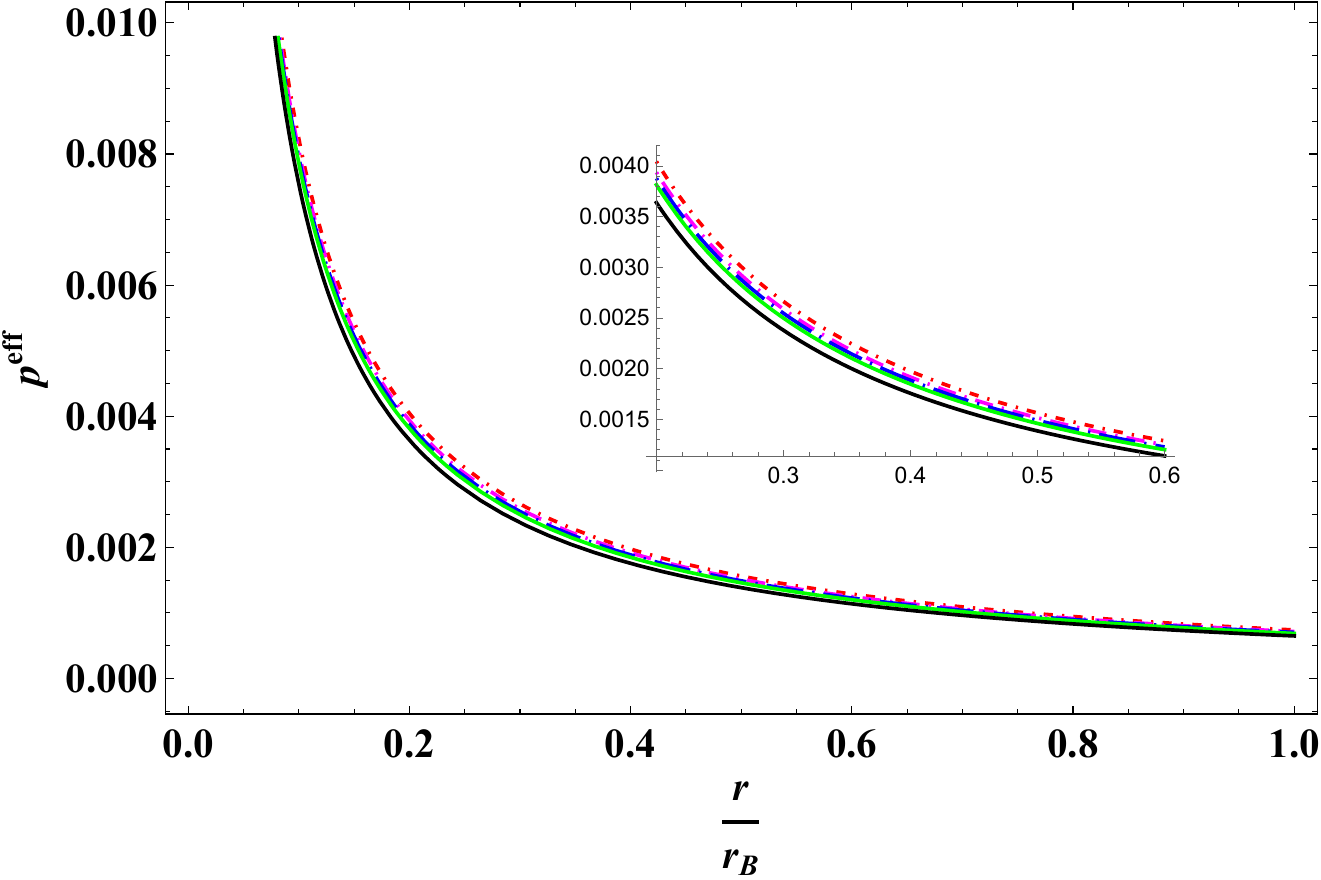}
      \caption{Behaviour of pressure and matter density ($\text{km}^{-2}$)for model I (upper panel) and model II (lower panel). Here, we consider $m=-2,n=0.02$ for model I and $m=2,n=-1$ for model-II.\label{pressure2}}
\end{figure*}
Moreover, the current investigation results in negative values for the derivatives of the energy density and pressure functions concerning the radial coordinate, denoted as $\frac{d\rho^{\text{eff}}}{dr}$ and $\frac{dp^{\text{eff}}}{dr}$, respectively. The presence of negative gradients in Fig.(\ref{dpdr}) indicates that the solutions we have discovered meet the physical requirement and are physically acceptable for both models.
\begin{figure*}[htbp]
    \includegraphics[width=8cm, height=5cm]{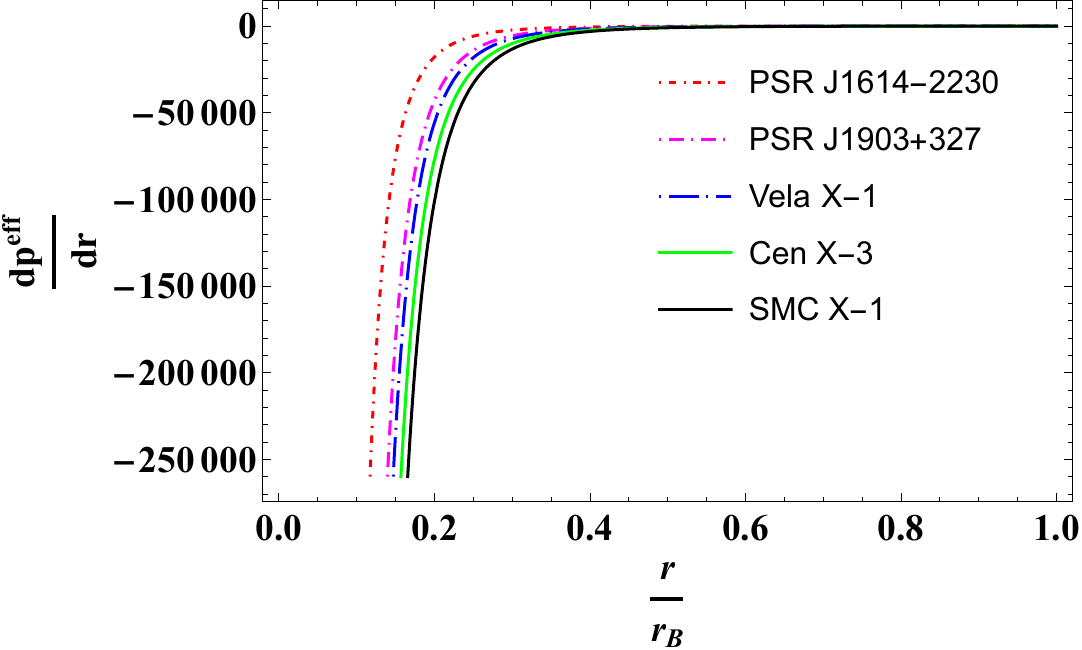}
    \includegraphics[width=8cm, height=5cm]{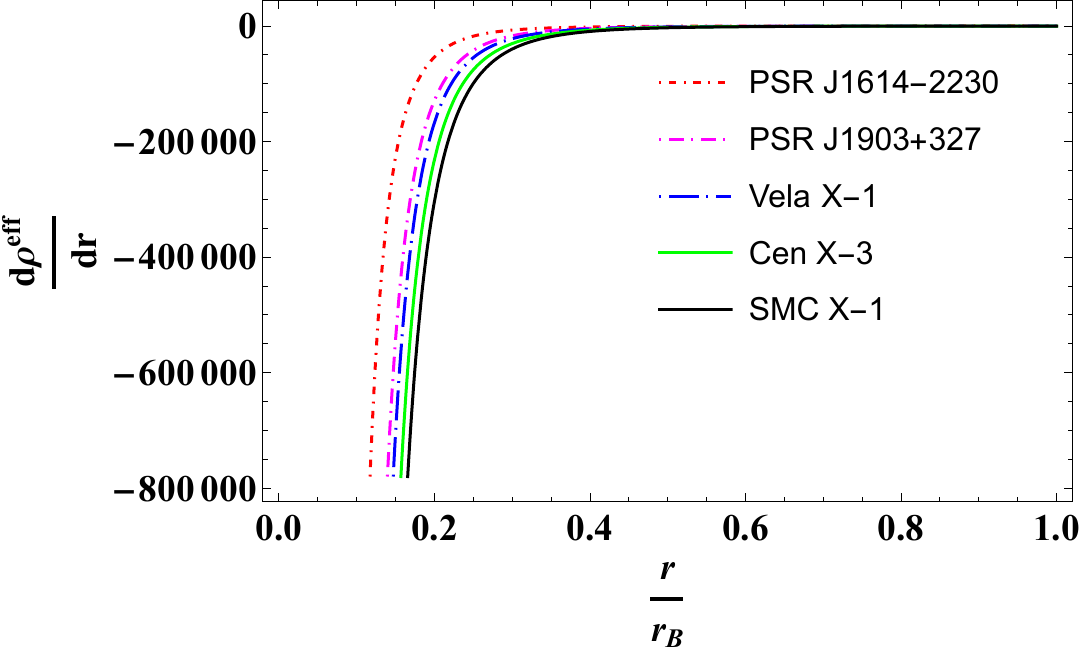}
    \includegraphics[width=8cm, height=5cm]{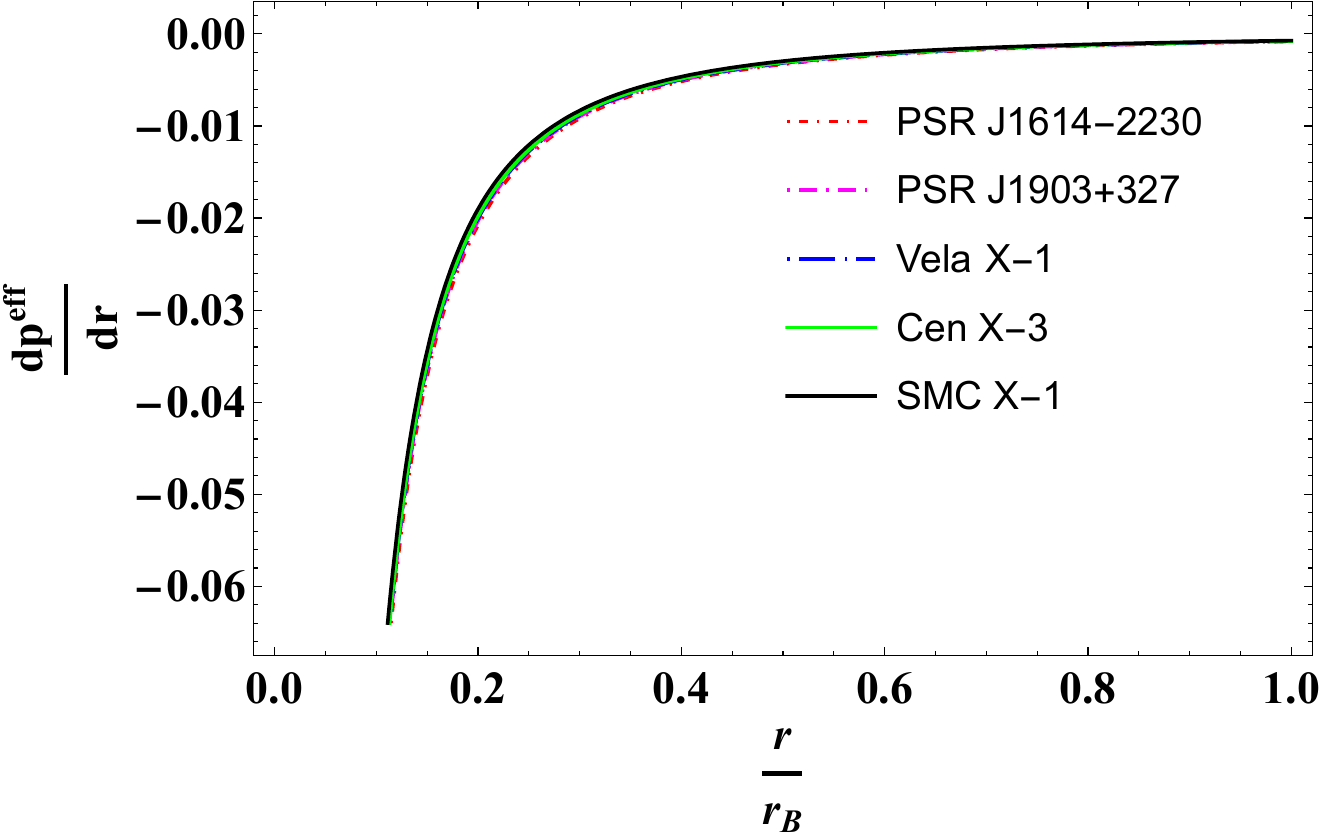}
    \includegraphics[width=8cm, height=5cm]{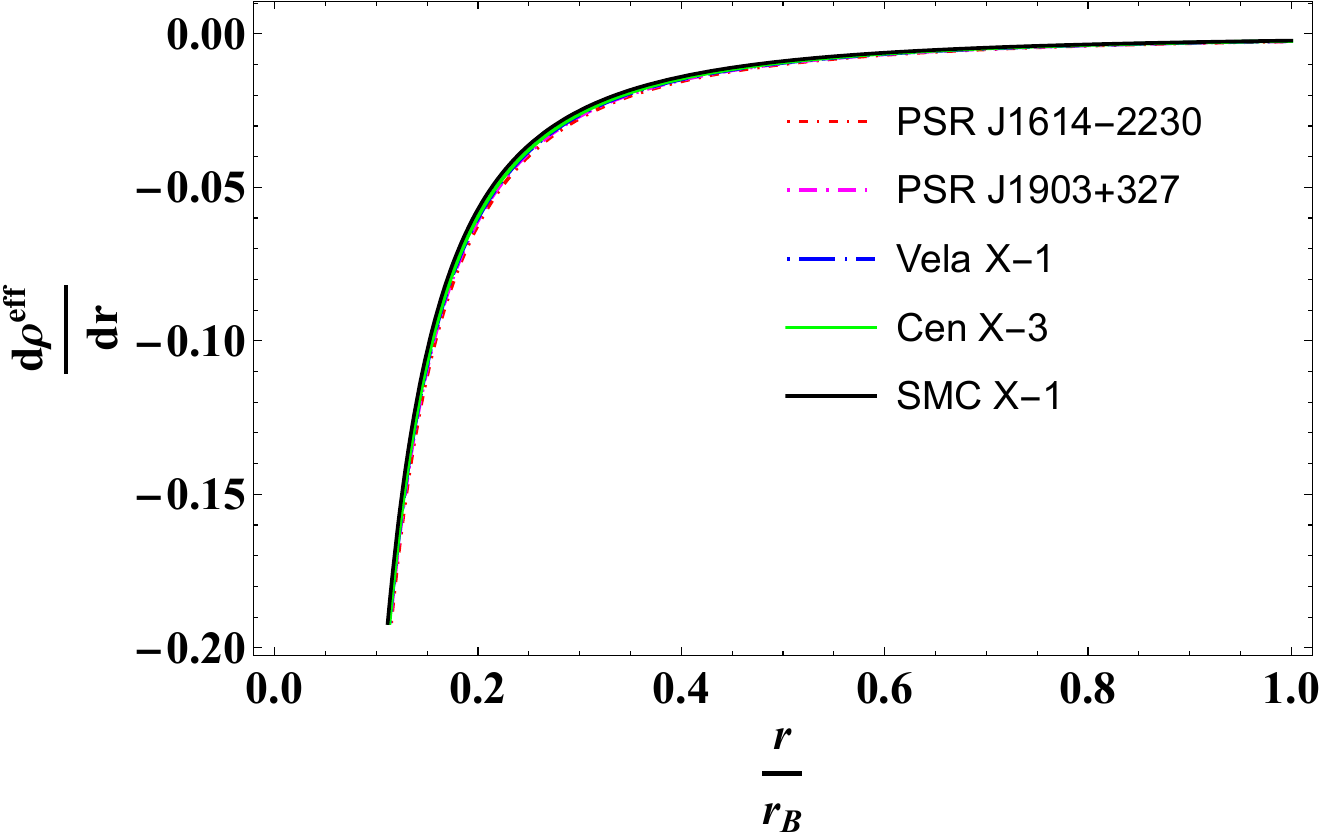}
      \caption{Behaviour of pressure gradient and matter density gradient for model I (upper panel) and model II (lower panel). Here, we consider $m=-2,n=0.02$ for model I and $m=2,n=-1$ for model-II.\label{dpdr}}
\end{figure*}
        \item \textbf{Energy condition :} The investigation of energy circumstances holds numerous important implications in the realms of GR and cosmology. The study of the Hawking-Penrose singularity theorems and the reliability of the second law of black hole thermodynamics can be facilitated by analyzing the energy circumstances \cite{HWA}. Relativistic cosmology explores many intriguing findings by utilizing energy conditions which are given below :
        \begin{itemize}
            \item Null energy condition ($NEC$) : $\rho^{\text{eff}}+p^{\text{eff}} \geq 0$
        \item Weak energy condition ( $W E C$ ): $\rho^{\text{eff}}\geq0, \rho^{\text{eff}}+p^{\text{eff}} \geq 0$.
        \item Strong energy condition ($SEC$) :$\rho^{\text{eff}}+3p^{\text{eff}}\geq0,\rho^{\text{eff}}+p^{\text{eff}}\geq 0$
        \item Dominant energy condition ($DEC$):$\rho \geq0 , \rho^{\text{eff}} \pm p^{\text{eff}} \geq 0$
        \end{itemize}
       From the Fig.(\ref{pressure2},\ref{energy}), it is evident that all the energy conditions are justified, indicating that the charged compact star under Bardeen space-time is feasible for both of our constructed models. The physical parameters like pressure and density exhibit a central singularity due to the use of conformal symmetries. Indeed, this formalism cannot overcome the core singularity in the physical parameters. In fact this is the only disadvantage of this conformal motion treatment which can be seen in the earlier studies in\cite{C1,C4}.

\begin{figure*}[htbp]
    \includegraphics[width=8cm, height=5cm]{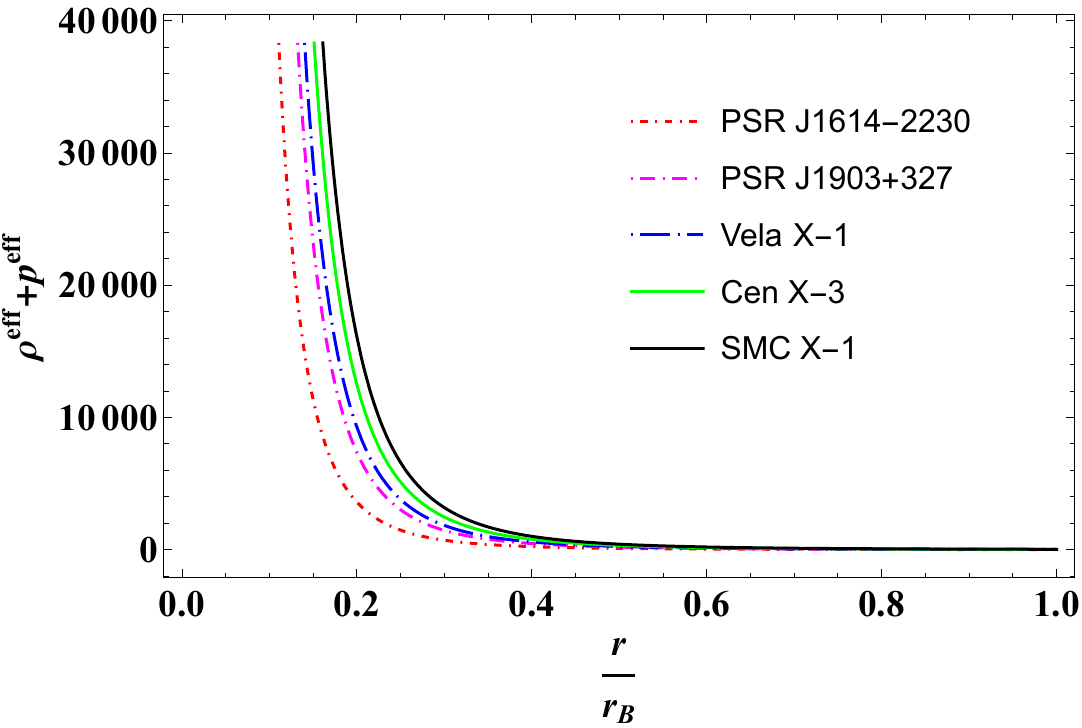}
    \includegraphics[width=8cm, height=5cm]{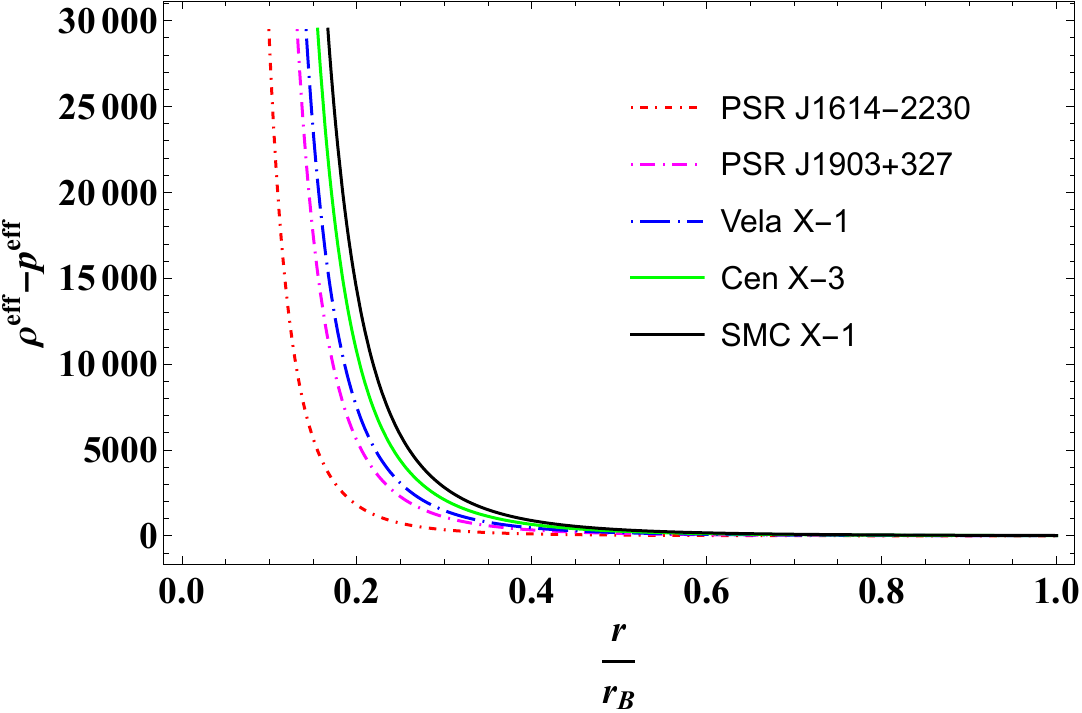}
    \includegraphics[width=8cm, height=5cm]{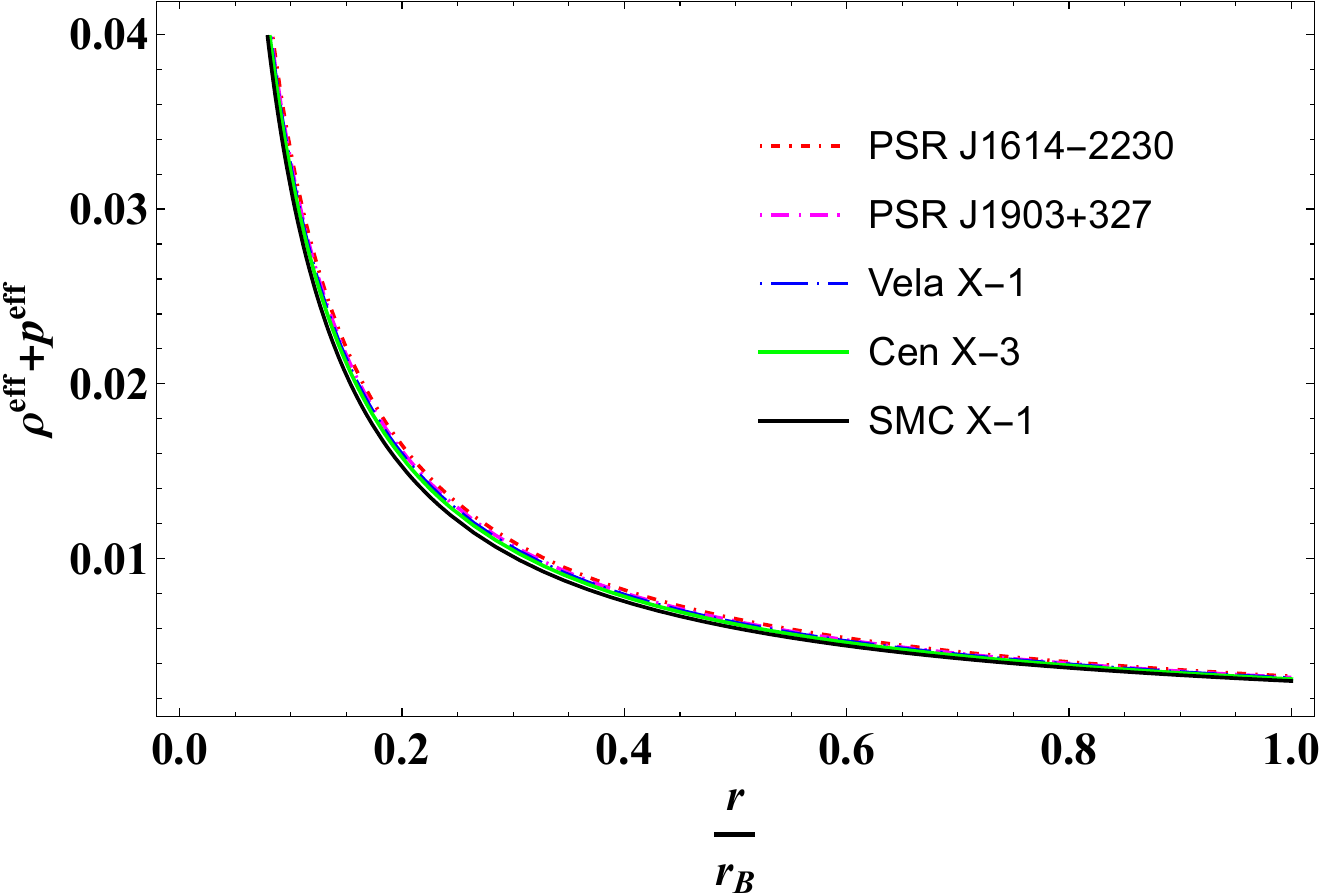}
    \includegraphics[width=8cm, height=5cm]{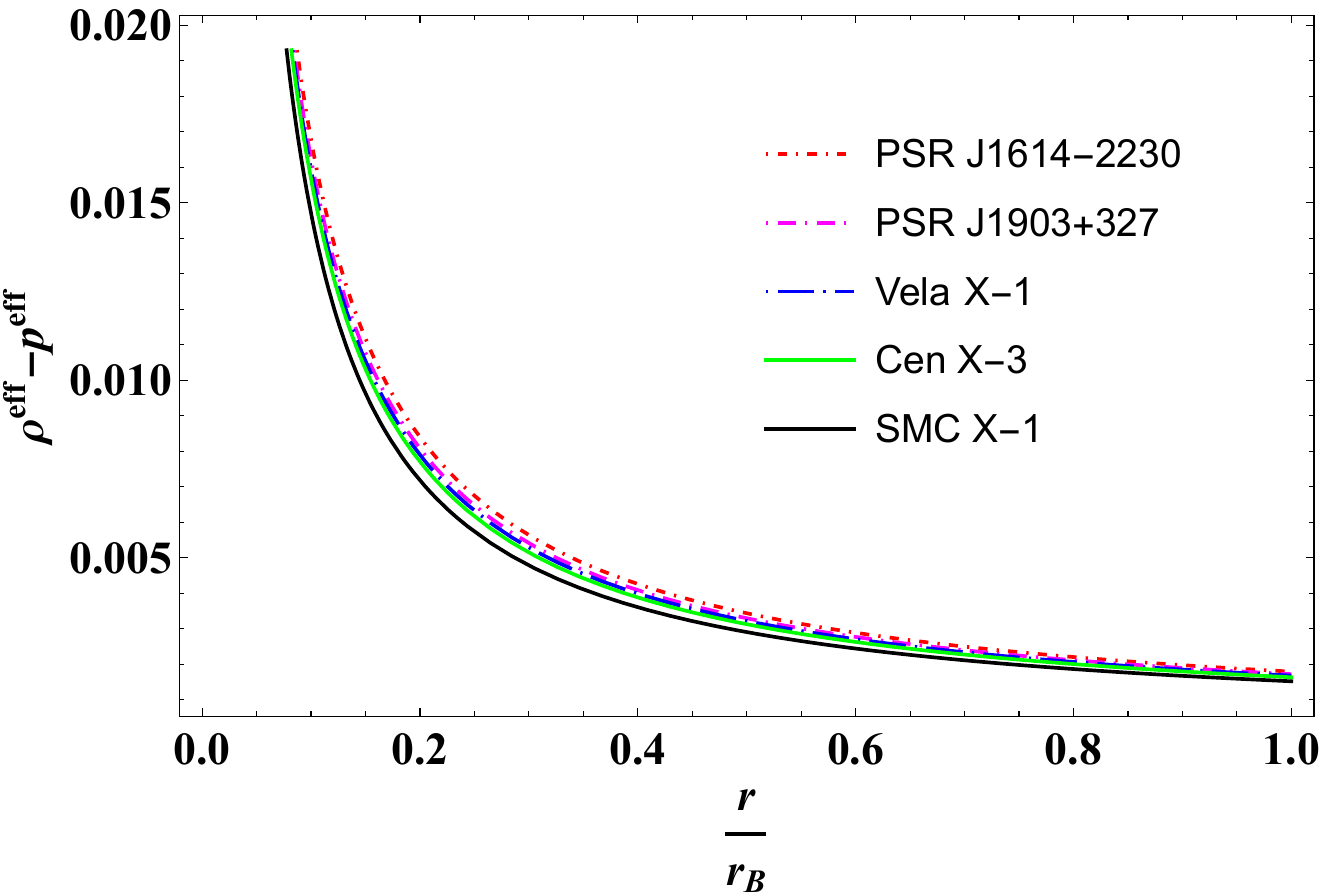}
      \caption{Behaviour of energy conditions for model I (upper panel) and model II (lower panel). Here, we consider $m=-2,n=0.02$ for model I and $m=2,n=-1$ for model-II.\label{energy}}
\end{figure*}
\item \textbf{Equation of State:}
An essential method for analyzing the relationship between matter density and pressure is to determine the equation of states. The formula for determining
the equation of state is $\omega=\frac{p^{\text{eff}}}{\rho^{\text{eff}}}$, $\omega$ is the state parameter. From the graphical analysis of Fig.\ref{pressure2} it is clear that the EoS parameter $\omega$ lies in the range $0<\omega<1$ (a preferable limit for compact stars), i.e., it lies inside the bounds of the radiation era.
\end{enumerate}
\section{Equilibrium and Stability analysis}\label{secv}

\begin{enumerate}
    \item \textbf{Causality requirement :} The causality condition must be maintained, which says that the magnitude of the speed of sound must be lower than the speed of light. In other words, the inequality $0\leq v^2=\frac {d p^{\text{eff}}}{d\rho^{\text{eff}}}\leq 1$ must be satisfied. In this investigation, we determined the square of the speed of sound $v^2=\frac{1}{3}$ maintains the above criterion for stability for both the case of model I and model II.
    \item \textbf{Relativistic Adiabatic Index:} The stability of a compact star can be discussed through another physical parameter, namely the adiabatic index. The adiabatic index, which is defined as $\Gamma_r$, can be expressed as,
    \begin{equation}
        \Gamma_r = \frac{\rho^{\text{eff}}+p^{\text{eff}}}{p^{\text{eff}}}\frac {d p^{\text{eff}}}{d\rho^{\text{eff}}}.
    \end{equation}
    This significant parameter encompasses two conditions described by Hillebrandt and Steinmetz \cite{z1}. Given this requirement, if the value of $\Gamma_r$ is more than $4/3$, it indicates the stability of a compact star. However, $\Gamma_r$ was supposed to be an unstable sphere with a value less than $4/3$. From Fig.(\ref{adb}), it is clear that all the curve for different compact stars maintains the range $\Gamma_r>\frac{4}{3}$. Through this study, the result $\Gamma_r>\frac{4}{3}$ proves that solutions with conformal motion and the Bardeen model as the star's outer space-time, meet the stability requirements based on the Adiabatic index.
\begin{figure*}[htbp]
    \includegraphics[width=8cm, height=5cm]{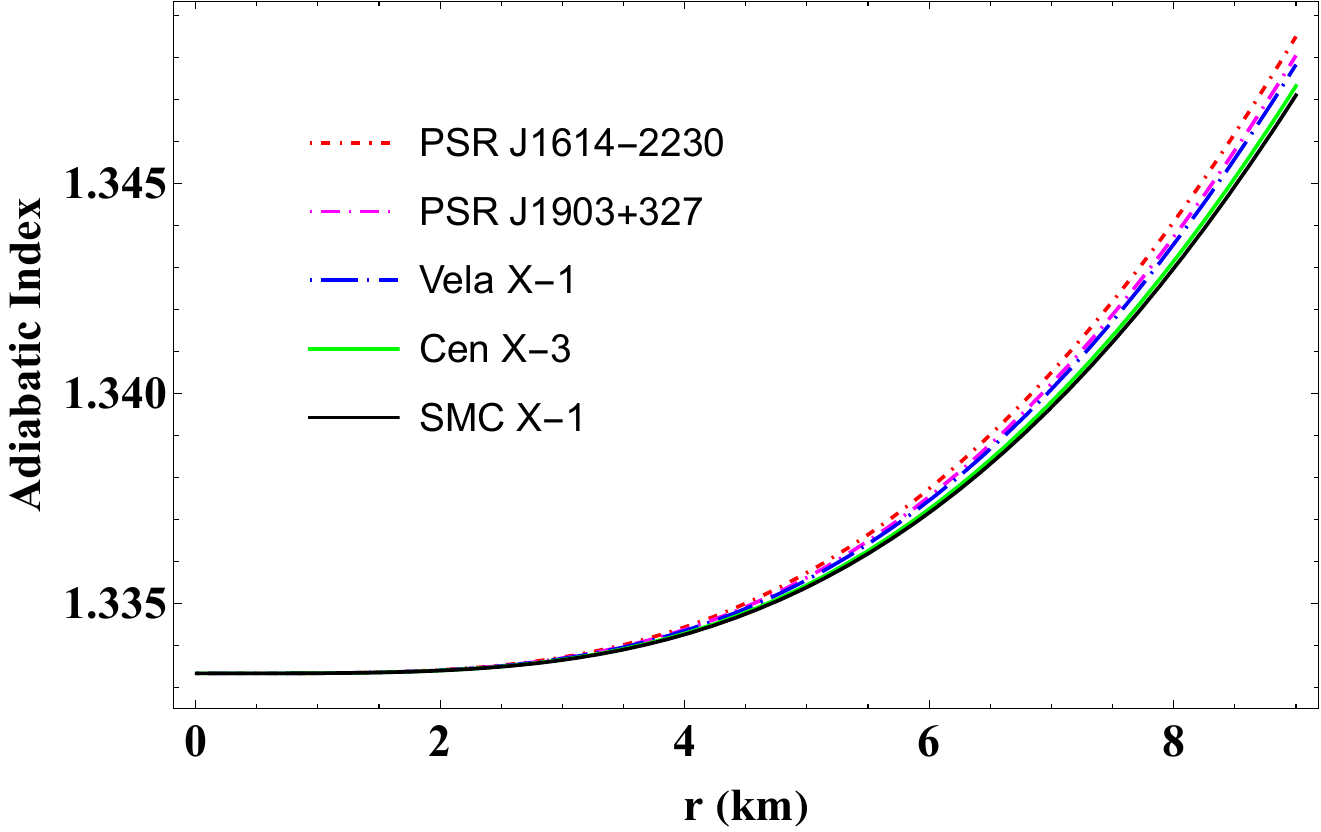}
    \includegraphics[width=8cm, height=5cm]{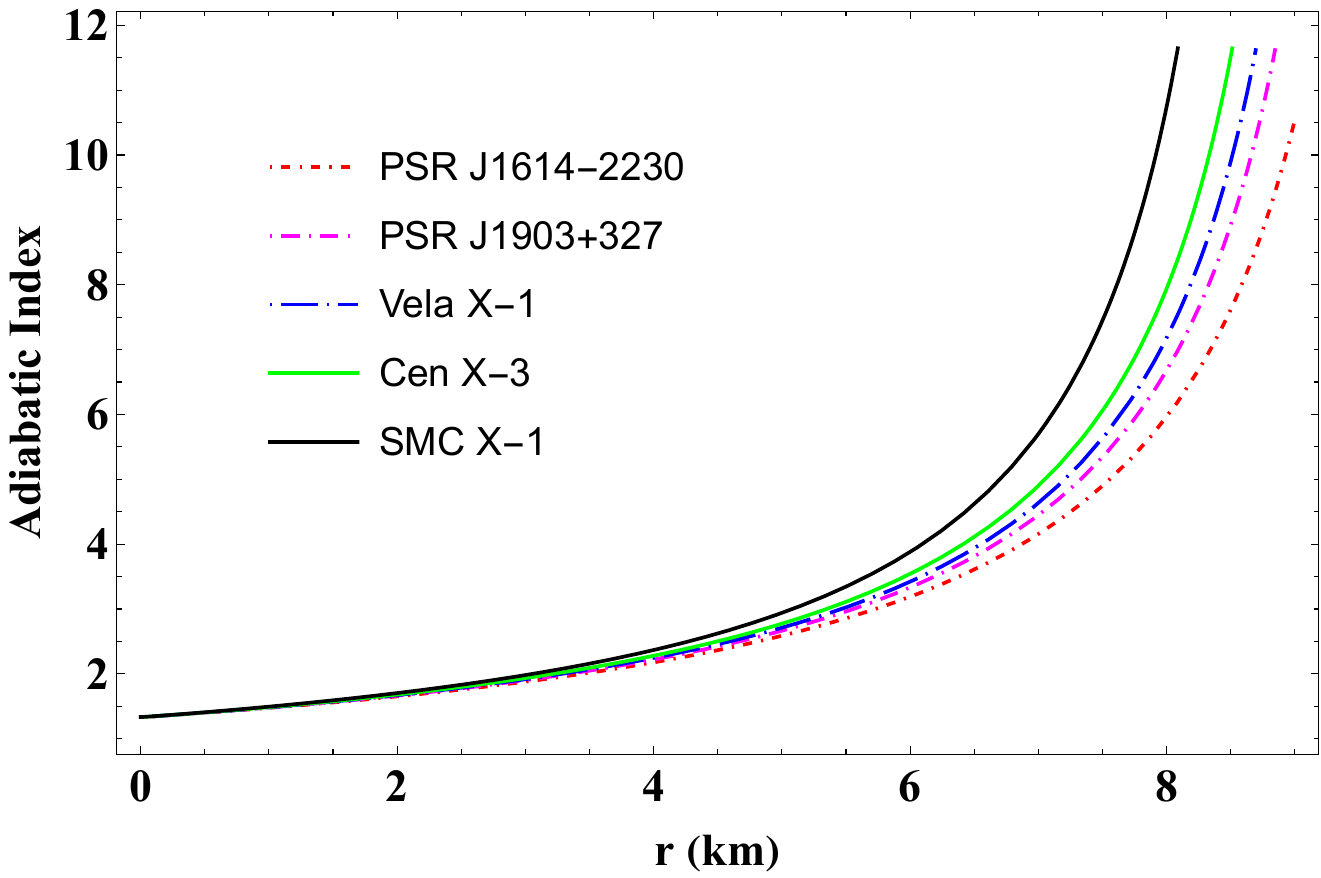}
      \caption{Behaviour of adiabatic index for model I (left panel) and model II (right panel).  Here, we consider $m=-2,n=0.02$ for model I and $m=2,n=-1$ for model-II. \label{adb}}
\end{figure*}
    \item \textbf{Equilibrium Conditions:} This subsection examines the equilibrium conditions for studying the current charged star structure in the background of Bardeen geometry, admitting the conformal motions. It is essential to note that the Tolman-Oppenheimer-Volkoff (TOV) equation is crucial for analyzing the equilibrium conditions of a stellar structure. A static spherically symmetric object in static gravitational equilibrium has its structure primarily constrained by the TOV equation. It is particularly intriguing to see the behavior of gravitational and other fluid forces as the electrostatic repulsion near the star's border increases. The explicit form of the different forces is given below.
    \begin{itemize}
        \item Gravitational forces :
        $F_g=-\frac{\nu^{\prime}}{2}(\rho^{\text{eff}}+p^{\text{eff}})$,
        \item Hydrostatic force :
        $F_h=-\frac{dp^{\text{eff}}}{dr}$,
        \item Electric force :
        $F_e=\sigma E e^{\frac{\lambda}{2}}$.
    \end{itemize}
     By combining, the TOV equation can be written as,
    \begin{eqnarray}
        F_g+F_h+F_e=0.
    \end{eqnarray}
    
    \begin{figure*}[htbp]
    \includegraphics[width=8cm, height=5cm]{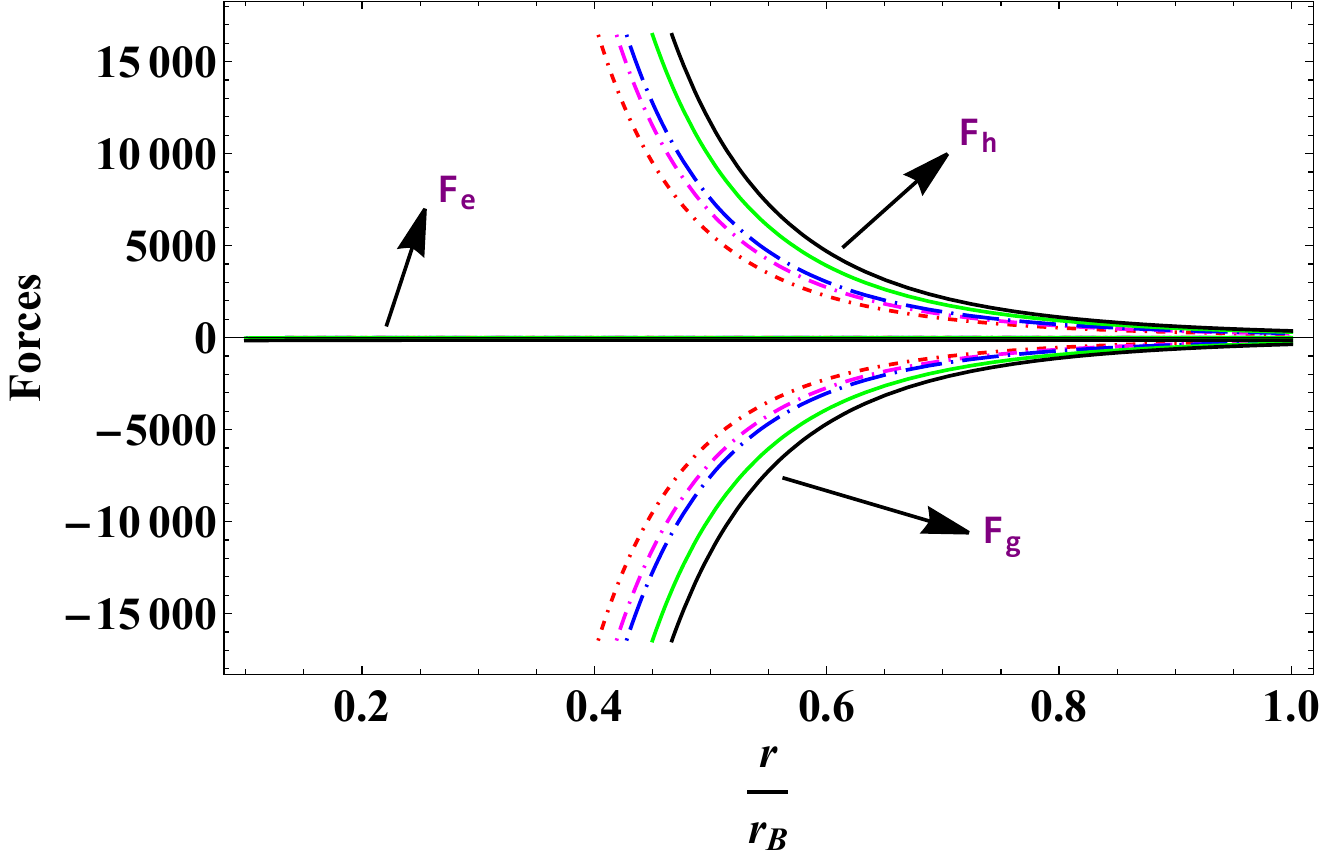}
    \includegraphics[width=8cm, height=5cm]{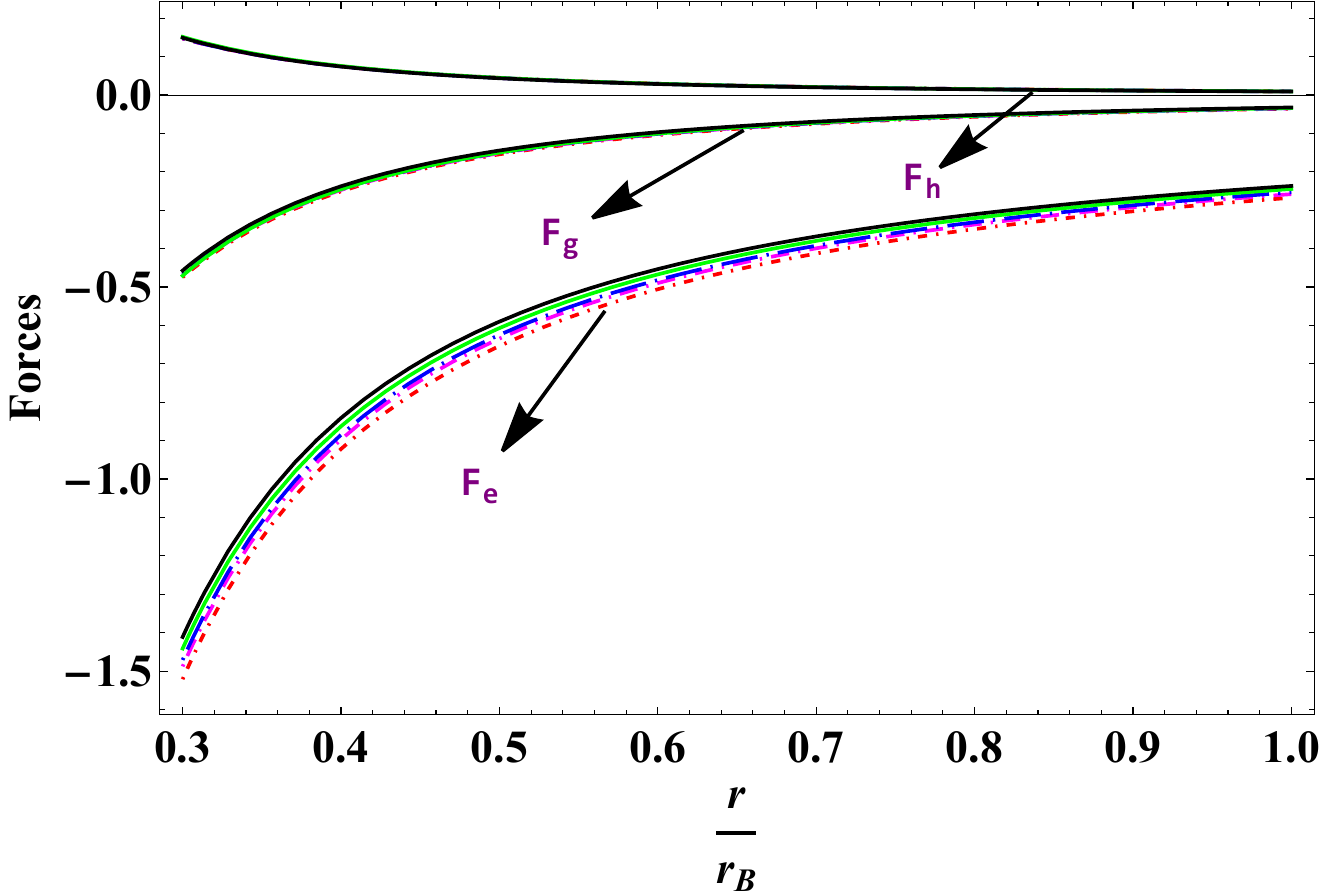}
      \caption{Behaviour of different forces for model I (left panel) and model II (right panel). The different colors represents,  PSR J1614-2230($\textcolor{red}\star$), PSR J1903+327 ($\textcolor{magenta}\star$), Vela X-1 ($\textcolor{blue}\star$), Cen X-3 ($\textcolor{green}\star$), and SMC X-1 ($\textcolor{black}\star$).  Here, we consider $m=-2,n=0.02$ for model I and $m=2,n=-1$ for model-II.\label{Forces}}
\end{figure*}
We have graphically analyzed all the different forces for various compact objects like PSR J1614-2230, PSR J1903+327, Vela X-1, Cen X-3, and SMC X-1, by utilizing different observational data on their mass and radius in Fig.\ref{Forces}. In model I, it is noteworthy that the gravitational force and hydrostatic force are in equilibrium, and the electric force remains constant along the x-axis. However, in the case of model II, while the gravitational force and hydrostatic force balance out each other, but the behavior of the electric force can lead to instability in the stellar model. Thus, we can see how these forces balance out using Bardeen geometry and conformal motion with electric charge. This shows that our constructed power-law model is more stable and physically acceptable than the linear model.

\item \textbf{Andreasson's Limit :} Prior research has determined that in the case of a black hole, collapse consistently occurs at a critical radius $R_c$ beyond the outer horizon when $Q<\mathcal{M}$. As the value of $Q$ approaches $M$, this critical radius approaches the event horizon \cite{y1}. In the case of a non-charged object, the situation is comparable to the one described by the Buchdahl inequality \cite{y2}, which states that collapse will occur when the radius $r_b<9\mathcal{M}/4$. The mass function for charged compact stars must satisfy Andreasson's limit \cite{y3} $\sqrt{\mathcal{M}}<\sqrt{\frac{r_b}{3}}+\sqrt{\frac{r_b}{9}+\frac{q^2}{3r_b}}$, which is given in table-\ref{table2} for both of the models.
\begin{table}
\caption{The estimated values of mass and Andreasson's limit}\label{table2}
\centering
  \begin{tabular}{@{}ccccccccccccc@{}}
            \hline\hline
             &   Model-I & \\
            \hline
             Stars & $\sqrt{\mathcal{M}}$ & $\frac{\sqrt{r_b}}{3}+\sqrt{\frac{r_b}{9}+\frac{q^2}{3r_b}}$\\
             \hline
            PSR J164-2230\cite{s1} & $1.40357$ & $3.64223$ \\
            PSR J1903+327\cite{s2} & $1.33041$ & $3.55699$ \\
            Vela X-1\cite{s3} & $1.29112$ & $3.50867$\\
            Cen X-3\cite{s3} & $1.22066$ & $3.46229$ \\
            SMC X-1\cite{s3} & $1.13578$ &$3.30119$\\
            \hline\hline
             &   Model-II & \\
            \hline
             Stars & $\sqrt{\mathcal{M}}$ & $\frac{\sqrt{r_b}}{3}+\sqrt{\frac{r_b}{9}+\frac{q^2}{3r_b}}$\\
             \hline
            PSR J164-2230\cite{s1} & $1.40357$ & $13.8682 $ \\
            PSR J1903+327\cite{s2} & $1.33041$ & $13.6346 $ \\
            Vela X-1\cite{s3} & $1.29112$ & $13.5013$\\
            Cen X-3\cite{s3} & $1.22066$ & $13.4968 $ \\
            SMC X-1\cite{s3} & $1.13578$ &$12.9331$\\
        \end{tabular}
\end{table}
\end{enumerate}

\section{Comparison with Reissner-Nordstrom case}\label{vi}

\begin{figure*}[htbp]
    \includegraphics[width=8cm, height=5cm]{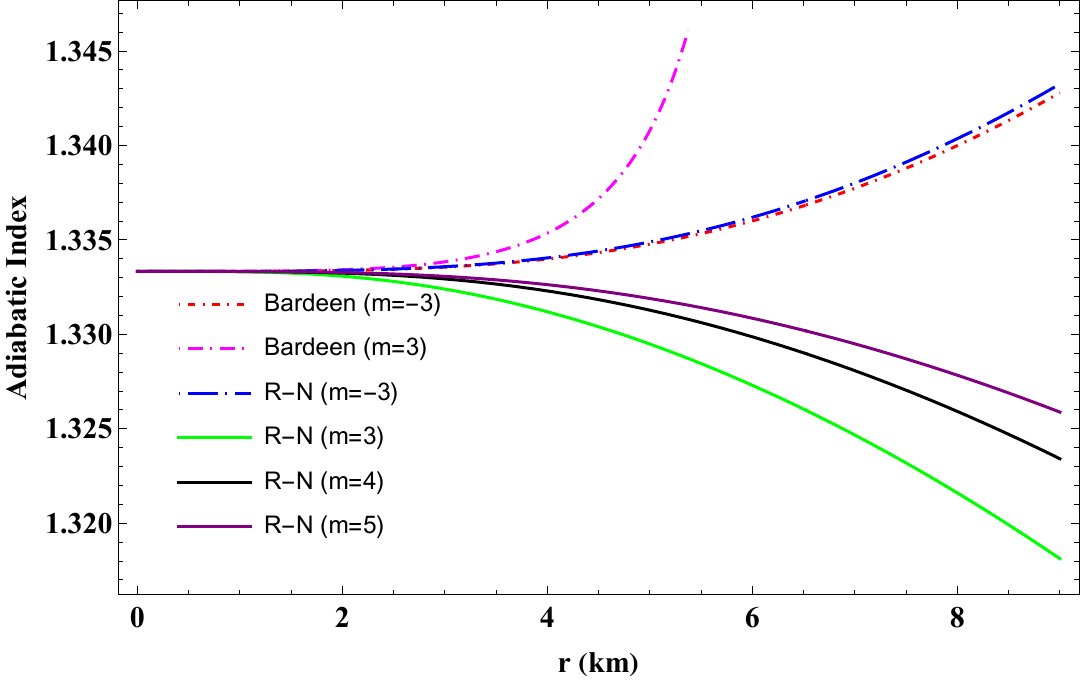}
    \includegraphics[width=8cm, height=5cm]{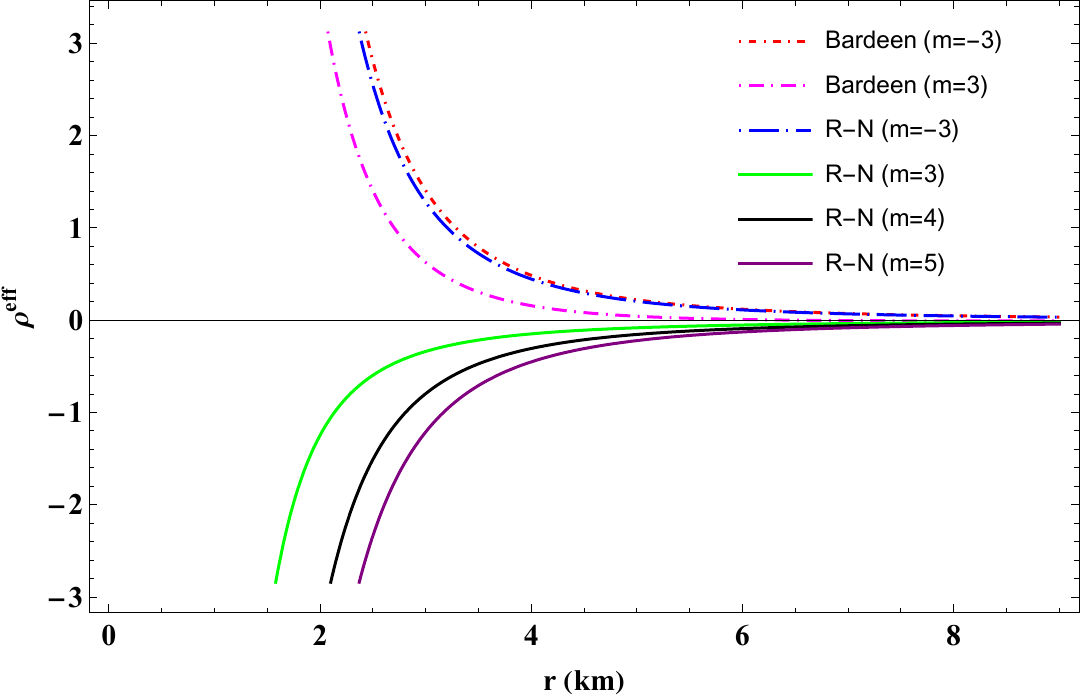}
      \caption{Comparison between the Bardeen and R-N space-time. \label{compare}}
\end{figure*}

Here, we present a brief overview of the scenario with Reissner-Nordstrom (R-N) spacetime as the external geometry for the matching condition. Additionally, this will allow us to make comparisons with our study. The R-N space-time is defined by:
\begin{eqnarray}
ds^2 &=& -\big(1-\frac{2 \mathcal{M}}{r}+\frac{ \mathcal{Q}^{2}}{r^{2}}\big) dt^2 +\big(1-\frac{2 \mathcal{M}}{r}+\frac{ \mathcal{Q}^{2}}{r^{2}}\big)^{-1}\nonumber\\ &&\hspace{-0.5cm}+r^2 (d\theta^2+sin^2 \theta d\phi^2).
\end{eqnarray}
Similarly, as in the previous case, we implement the continuity equation to get the constant of our proposed model I, and we have shown those comparative studies in Fig (\ref{compare}).\\
In this comparative study, we have examined the energy density and adiabatic index, which play an important role in studying a stellar model's physical behavior and stability. We have varied the model parameter in a wide range and observed that, for the Bardeen space-time, the positive behavior of energy density and its increasing behavior towards the star's core, but for the R-N model, as the model parameter $m$ increases, we get the negative energy density and it gradually decreases towards $-\infty$ which is not feasible in the regime of astrophysical object. Besides, from the second panel of Fig.\ref{compare}, readers can observe that our proposed model-I with the Bardeen space-time gives the value $\Gamma_r>\frac{4}{3}$ for a wide range of $m$. But, in the case of R-N space-time, as $m$ increases, it maintains the adiabatic index limit $\Gamma_r>\frac{4}{3}$ for a certain range of radius, particularly in the core region, but as it approaches towards the outer surface region, it doesn't maintain the above-mentioned limit, which doesn't support proving a viable model. However, our constructed model II doesn't show any significant differences between the Bardeen and R-N space-time.

\section{Conclusion}\label{vii}

This study examines the solutions of field equations for investigating compact stars in the context of $f(Q)$ gravity. To achieve this aim, we have considered a perfect fluid as the source of matter, which also possesses an electric charge. In this analysis, we select the MIT Bag model equation of state (EoS) to describe the link between pressure and energy density. In addition, CKV's are utilized to examine the suitable configurations for gravitational metric coefficients. The MIT Bag model EoS is important in formulating a differential equation by utilizing field equations and solutions via conformal motion. This offers us a foundational mathematical basis for subsequent analysis. Here, we have studied two models of conformal factor $\phi(r)$. In model I, we have derived the solution of $\phi(r)$ by incorporating a functional form $\phi(r)=I\sqrt{\psi(r)}$ and using the field equation. We must note that we do not consider the integration constant $C$ zero when solving the differential equation. Our current study of compact stars relies significantly on this integration constant. Whereas the second model we have studied by taking conformal factor $\phi(r)$ simply as a linear function of $r$. In addition, we select the Bardeen model to represent the outside spacetime, which allows us to impose the boundary requirements. With remarkable precision, the Bardeen model presents the information originally and interestingly. Also, in a specific case of non-linear electrodynamics, the Bardeen solution can be seen as a gravitationally collapsed magnetic monopole. Here is an overview of the current work's qualitative analysis:

\begin{itemize}
    \item \textbf{Metric potential:} For the model I, the metric potential gives the positive, bounded, and finite values throughout the stellar configuration. Also, it avoids any kind of stellar singularity in the region $0<r<r_b$. Besides, our constructed model II shows some unacceptable properties at the core, as $e^{\lambda}\to \infty$ when $r\to0$. Except for the core, it gives positive, finite, and bounded values up to the stellar boundary. So, from here, we can conclude that for studying the Bardeen star in $f(Q)$ gravity through the conformal motion, it is better to consider the power-law form of the conformal factor than the linear form. 
    \item \textbf{Nature of physical quantity:} The changes in energy density and pressure functions are displayed in Fig.\ref{pressure2}. The energy density and pressure profiles exhibit realistic behavior, with the exception of an unavoidable central singularity. The value of $\rho$ is observed to be positive, gradually dropping to a minimum at $r=r_b$, while the pressure tends to approach zero at the star's edge. The whole analysis is done by considering some compact objects for different observational data of their radius and mass. Apart from that, the negative pressure gradient and density gradient in Fig(\ref{dpdr}) confirm the stability of our constructed model.
    \item \textbf{Energy condition:} From Fig.(\ref{pressure2},\ref{energy}), it is evident that all the energy conditions are justified, indicating that the charged compact star under Bardeen space-time is feasible for both of our constructed models.
    \item \textbf{Stability analysis:} The stability analysis plays an important role in studying the compact object in the background of GR or any modified gravity. The causality condition is one of the important features to check the stability. It says that the speed of sound inside the star must be subluminal. For both of our constructed models, we obtained $v^2=\frac{1}{3}<1$. Apart from that, from Fig.\ref{adb}, it is clear that the value of the adiabatic index $\Gamma>\frac{4}{3}$. Therefore, the EoS of the neutron star with the quark matter meets the stability criterion. For a charged compact star, the mass function must satisfy Andreasson's limit, which is shown in table \ref{table2}. For model I and model II, one can see that $\sqrt{\mathcal{M}}<\frac{\sqrt{r_b}}{3}+\sqrt{\frac{r_b}{9}+\frac{q^2}{3r_b}}$ which satisfy the aforementioned Andreasson's limit.
    But, in model II, the value of $\frac{\sqrt{r_b}}{3}+\sqrt{\frac{r_b}{9}+\frac{q^2}{3r_b}}$ is slightly higher than expected, as previously explored in the work of Bardeen Star. However, in the study of equilibrium conditions via the TOV equation, we have observed clearly how all the three forces $F_g, F_h, F_e$ balance out each other for our constructed model I. Whereas in the case of model II, while the gravitational force and hydrostatic force balance out each other, the behavior of the electric force can lead to instability in the stellar equilibrium.
\end{itemize}

Additionally, we presented a brief overview of the scenario where the matching condition is satisfied using R-N spacetime as the exterior geometry. Through this, we can make some comparisons and draw conclusions about our study. For comparison with GR \cite{C1}, we would like to describe that, here in his study, we have extended the work up to considering two models of conformal factor and compared our obtained solution by using Bardeen's geometry and R-N geometry for model-I. In the first step, we have examined the density behavior for both space-time by taking the same values of the model parameters $m$ and $n$ in Fig.\ref{compare}. We observe that the energy density exhibits positive behavior within Bardeen space-time and increases as it approaches the star's core. However, in the R-N model, as the model parameter $m$ increases, the energy density becomes negative and gradually decreases towards $\infty$, which is not feasible within a compact star. In the second comparison, from the second panel of Fig.\ref{compare}, one can see that Bardeen space-time yields $\Gamma_r>\frac{4}{3}$ for a large range of $m$. But, in R-N space-time, as $m$ increases, the adiabatic index limit $\Gamma_r>\frac{4}{3}$ is maintained for a certain radius range but not for the outer surface region. This prevents the proof of a viable model. The above comparison is made for the power law model by considering the compact star PSR J1903+327. However, our constructed linear model shows no major differences between Bardeen and R-N space-time. 

To summarize, we can look into a stable and possible structure for a compact star with charge using the modified $f(Q)$ gravity model with Bardeen's black hole geometry and conformal motion. By employing two models of conformal factors, we have shown that the power-law model gives a better result than our constructed linear model. Conformal symmetries are useful for building the mathematical formulation of a physical solution in compact star research, but they have a major flaw: they have a singularity in the center. Except for the core singularity, the computed results utilizing Bardeen geometry as an outer space-time are inherently well-behaved.  

\textbf{Data availability} There are no new data associated with this article.

\acknowledgments PKS  acknowledges the National Board for Higher Mathematics (NBHM) under the Department of Atomic Energy (DAE), Govt. of India for financial support to carry out the Research project No.: 02011/3/2022 NBHM(R.P.)/R \& D II/2152 Dt.14.02.2022 and IUCAA, Pune, India for providing support through the visiting Associateship program.

\end{document}